\documentclass[a4paper,fleqn,usenatbib,onecolumn]{mnras}

\usepackage{newtxtext,newtxmath}

\usepackage[T1]{fontenc}
\usepackage{ae,aecompl}

\usepackage{graphicx}	
\usepackage{amsmath}	
\usepackage{amssymb}	

\usepackage[OT2,T1]{fontenc}
\DeclareSymbolFont{cyrletters}{OT2}{wncyr}{m}{n}
\DeclareMathSymbol{\Sha}{\mathalpha}{cyrletters}{"58}

\newcommand{\Conv}{\mathop{\scalebox{1.5}{\raisebox{-0.2ex}{$\ast$}}}}%

\title[Noodle Model for Scintillation Arcs]{Noodle Model for Scintillation Arcs}

\author[C. R. Gwinn]{
Carl R. Gwinn$^{1}$\thanks{E-mail: cgwinn@ucsb.edu } \\
$^{1}$Physics Department, University of California, Santa Barbara 93106, USA\\
}

\date{Accepted XXX. Received YYY; in original form ZZZ}

\pubyear{2018}

\begin{document}
\label{firstpage}
\pagerange{\pageref{firstpage}--\pageref{lastpage}}
\maketitle

\begin{abstract}
I show that narrow, parallel strips of phase-changing material, or ``noodles,'' generically produce parabolic structures in the delay-rate domain.
Such structures are observed as ``scintillation arcs'' for many pulsars.
The model assumes the strips have widths of a few Fresnel zones or less, and are much longer than they are wide.
I use the Kirchhoff integral to find the scattered field.
Along the strips, integration leads to a stationary-phase point where the strip is closest to the line of sight.
Across the strip, the integral leads to a 1D Fourier transform.
In the limit of narrow bandwidth and short integration time, the integral reproduces the observed scintillation arcs
and secondary arclets.
The set of scattered paths follows the pulsar as it moves.
Cohorts of noodles parallel to different axes produce multiple arcs, as often observed.
A single strip canted with respect to the rest produces features off the main arc.
I present calculations for unrestricted frequency ranges and integration times;
behavior of the arcs matches that observed, and can blur the arcs.
Physically, the noodles may correspond to filaments or sheets of over- or under-dense plasma,
with a normal perpendicular to the line of sight.
The noodles may lie along parallel magnetic field lines that carry density fluctuations, perhaps in reconnection sheets.
If so, observations of scintillation arcs would allow visualization of magnetic fields in reconnection regions.
\end{abstract}




\section{Introduction}\label{sec:Intro}

\subsection{Background}

Scintillation arcs are a remarkable phenomenon in interstellar scintillation of pulsars at decimeter and meter wavelengths.
They indicate the presence of extremely compact, sparsely distributed structures that are concentrated in thin screens, 
and produce angular deflections of radio waves by many milliarcseconds.
In general, scintillation appears as variations of intensity, as scattered paths reinforce or cancel.
The dynamic spectrum, a time series of spectra gathered sequentially in time, represents the intensity of the electric field $I$ as a function of observing frequency $\nu$ and time $t$.

\begin{figure}
\centering
\includegraphics[width=0.38\textwidth]{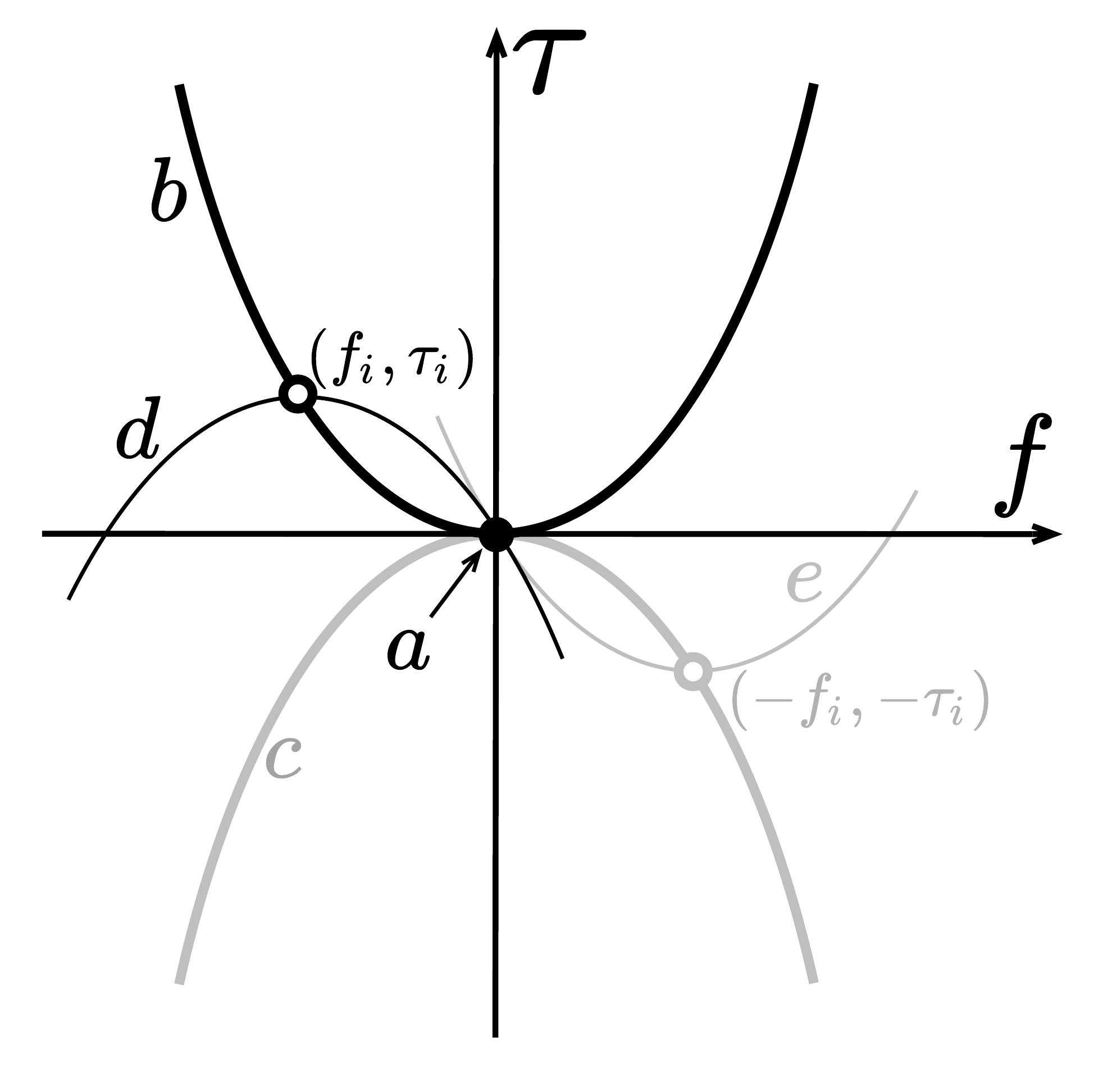}
\caption{Schematic depiction of scintillation arcs in the rate-delay plane $(f,\tau)$. 
The central maximum is the point at $a$ 
(the first term in Equation\ \ref{eq:MultiStripIntensityGeneral}, from the direct path);
the primary arcs are $b$ and its inversion $c$ (from the second and third terms);
and examples of secondary arclets are $d$ and $e$, the fourth and fifth terms.
\label{fig:reference_arcs}}
\end{figure}

First described by \citet{2001ApJ...549L..97S}, 
scintillation arcs appear in the secondary spectrum.
The secondary spectrum is
the square modulus of the Fourier transform of the dynamic spectrum to the domain of delay $\tau$ (Fourier conjugate to offset from center frequency $\Delta\nu = \nu - \nu_0$)
and rate $f$ (conjugate to time $t$).
Some authors refer to rate as Doppler frequency.
A maximum at the origin of the the secondary spectrum, $(\tau, f)=0$, represents the mean intensity of the source, averaged over the observed ranges of frequency and time.
Arcs extend from this maximum, along approximately parabolic paths $\tau= \pm a f^2$,
where $a$ is the curvature parameter, a constant.
Arcs appear for both positive and negative $\tau$: symmetrically as required for a power spectrum $I(\Delta \nu, t)$ (as for single-antenna observations),
or nearly symmetrically for a nearly real cross-power dynamic spectrum $V(\Delta \nu, t)$ (as for a baseline short compared with the lateral scale of scintillation).
Note that even for a short baseline, phases may extend over $2\pi$ in the Fourier-conjugate $(\tau, f)$ domain, as in Figure 1 of \citeauthor{2010ApJ...708..232B}
Particularly sensitive observations reveal secondary arclets, extending from points along each primary arc, and with the same curvature as the primary arc but in the opposite direction.
Figure\ \ref{fig:reference_arcs} displays a schematic view of a scintillation arc in the secondary spectrum.

Scintillation arcs have been observed for most nearby pulsars \citep{2007A&AT...26..517S}.
Even the power summed over all arcs and secondary arclets is much less than the power in the central
maximum \citep{2011ApJ...733...52G}.
Observations in several frequency bands at individual epochs show that 
the curvature parameter varies quadratically with observing frequency $\nu_0$:\ $a\propto \nu_0^2$ \citep{2003ApJ...599..457H}.
Arcs vary in strength from one observation to another;
in some cases, more than one arc is observed at one epoch \citep{2006ChJAS...6b.233P,2007ASPC..365..254S,2007A&AT...26..517S}.
Structures within the arcs evolve on timescales of months,
with individual identifiable features occasionally moving along the arcs, toward increasing $f$ \citep{2005ApJ...619L.171H}.
Interferometric observations have shown that the scintillation arcs arise from lines of sight separated in angle from the 
central maximum in the secondary spectrum, in an elongated distribution about a single axis \citep{2010ApJ...708..232B}.

Observations imply that the arcs are an interference phenomenon. Their weakness relative to the central peak indicate that the amplitude of the scattered field is small
relative to that of the unscattered field.
The wide span of the arcs in delay, far more than the inverse of the observing bandwidth, indicates that phase differences between scattered and unscattered fields amount to 
many turns of phase.
The identification of individual features suggests that the arcs arise from a set of entities $\{x_i\}$, mapped onto 
individual points of delay and rate $(\tau_i, f_i)$, via functions $\tau_i \propto a x_i^2$ and $f_i \propto x_i$.
The fact that interferometry detects phase differences along the arcs suggests that geometric phase,
from the offsets of scatterers from the line of sight, contributes to delay and rate.

In general, phase-coherent scattering can fall into ``strong'' and ``weak" extremes. In strong scattering the observer receives radiation from along multiple paths, with phases that differ by more than $2\pi$.
In weak scattering, phases along paths to the observer differ by less than $2\pi$.
Both the geometric path length and the optical path length can contribute to this phase, and both vary with frequency. The rapid change of arc properties with frequency, as evidenced by their large delay $\tau$ relative to the unscattered path, implies that the arcs result from strong scattering, in this sense.

Commonly, interstellar scattering is assumed to be ``optically thick'' in the sense that every path from the source intercepts the scattering material and is deflected to some degree \citep{CohenCronyn1974}. The deflection is drawn from a Gaussian or closely similar distribution.
This appears not to be the case for the scintillation arcs, at least in many cases:
only a small fraction of the radiation from the source suffers enough deflection to contribute to the arcs.
The rest resides in the nearly-undeflected radiation that contributes to the maximum at the origin in the secondary spectrum.
Thus, the arcs arise from strong, but ``optically thin,'' scattering.
In this sense, scintillation arcs are an extreme case of scattering by a Levy distribution,
where extreme but rare deflections dominate averages \citep{2003ApJ...584..791B,2003PhRvL..91m1101B,2005ApJ...624..213B}.

Traditionally, interstellar scattering has been treated as localized in a screen that is thin along the line of sight.
This picture includes many features of scattering material that is distributed along the line of sight.
The thin-screen picture also describes scintillation arcs well; models that include material extended along a large fraction of the line of sight tend to produce blurred arcs.

\subsection{Interpretations}

Interpretations of scintillation arcs have focused on discrete, compact scattering structures in the interstellar medium.
These compact structures deflect pulsar radiation to the observer from out of the line of sight \citep{2004MNRAS.354...43W}. 
The resulting paths interfere with the undeflected, primary path to produce interference fringes, with fringe phase that varies linearly with frequency and time.
Each such path produces a point in the secondary spectrum; the set of such paths forms the scintillation arc. 
In this interpretation, the paths remain stable over the time and frequency span of an observation,
and indeed over weeks, so that the motion of pulsar and observer shift features along the arcs, as \citeauthor{2003ApJ...599..457H} observed.
The stability of the arcs can be taken to suggest that the compact structures responsible for the scattering remain nearly fixed with changes in frequency and time.
Within this interpretation, a number of inversions have found the locations of the compact structures responsible for scattering on the sky, relative to the undeflected line of sight
\citep{2004MNRAS.354...43W,2010ApJ...708..232B, 2014MNRAS.440L..36P}.
The distribution of the compact scattering structures is highly anisotropic on the sky,
as found from inversions of single-dish and interferometric data.
The gradient of refractive index required to deflect a ray implies a large gradient of refractive index,
and hence a large column density of electrons, even given the small lateral scale, as I discuss quantitatively in a further paper (Paper II).

The physical nature and origin of the compact structures responsible for scintillation arcs has led to much discussion,
fueled by the inferences of small size and large column densities.
Among the suggestions are
evaporating concentrations of primordial hydrogen \citep{2013MNRAS.434.2814W}, 
quark strangelets \citep{2013PhLB..727..357P},
and filaments of gas from hot stars \citep{2017ApJ...843...15W}.
\citet{2014MNRAS.442.3338P} 
suggest that turbulence in reconnection sheets in the interstellar plasma, where those sheets happen to lie tangent to the line of sight, is responsible for the scattering.
A corrugated sheet can align with the line of sight at a number of places; each such location acts as a scatterer \citep{2016MNRAS.458.1289L}. 
The extension of the sheet along the line of sight spreads the required electron column over a longer distance,
reducing the required density fluctuation.
\citet{2018MNRAS.478..983S} 
propose a specific optical model for such a sheet, matching positions of subimages as measured in the splendid observations of scintillation arcs for pulsar B0834+06 by \citet{2010ApJ...708..232B}. 

\begin{figure}
\centering
\includegraphics[width=0.45\textwidth]{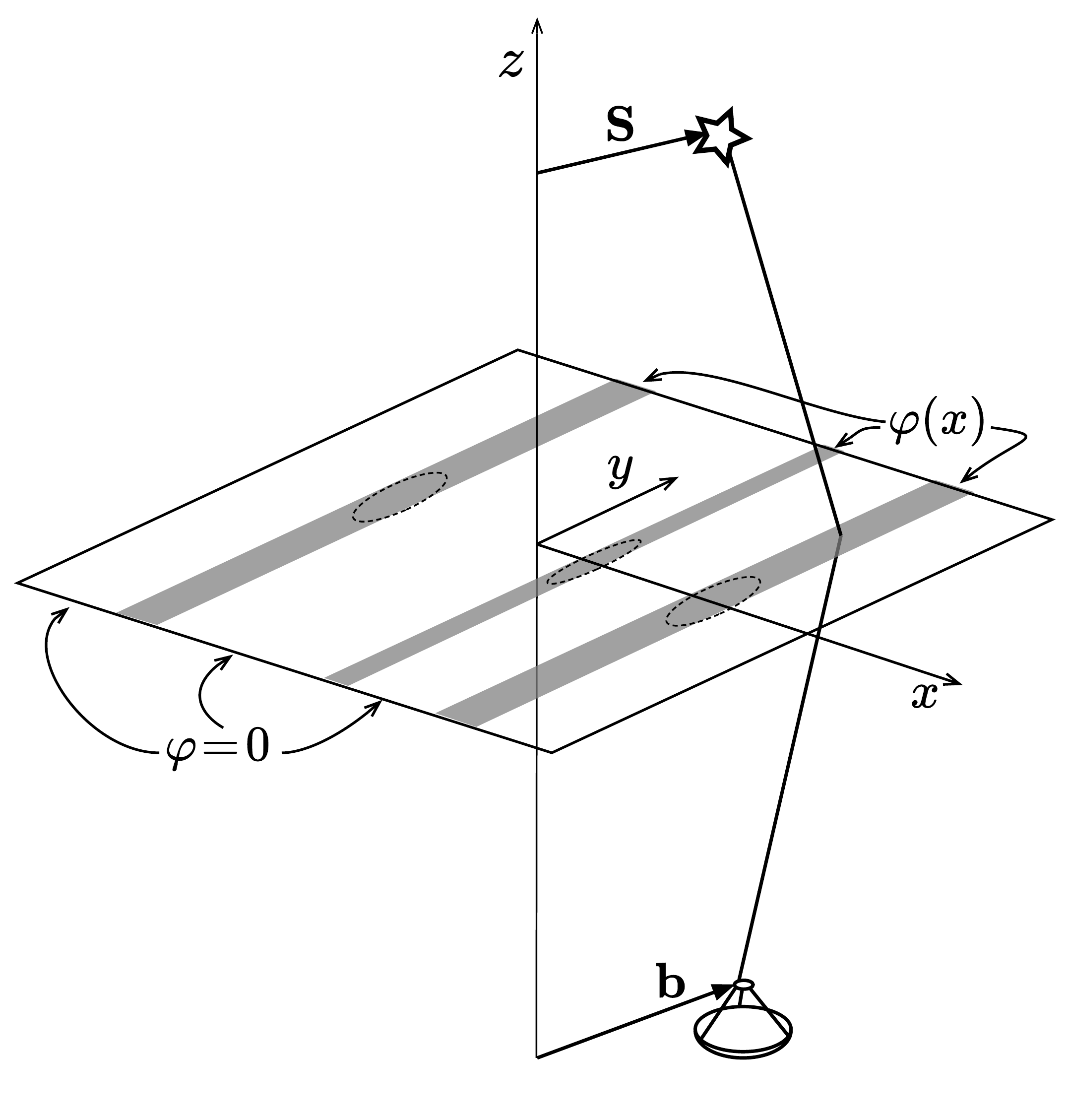}
\caption{Structure of ``noodle'' scatterers. The medium is uniform, except for narrow parallel strips in a thin plane perpendicular to the line of sight. The plasma phase may vary across the strips, but is uniform along them.  The dotted ellipses suggest contributions of the strips to the Kirchhoff integral.
\label{fig:basic_idea}}
\end{figure}

\subsection{Proposed Interpretation: Magnetic Noodles}\label{sec:PhysicalNature}

In this paper, I consider a simple picture in which variations in refractive index take the form of strips on a screen localized along the line of sight. Figure\ \ref{fig:basic_idea} illustrates the optics. 
Physically, these strips are ``noodles,'' either sheets or filaments, with a normal perpendicular to the line of sight.
Optically, they are equivalent to their projections onto the scattering screen, and thus to strips on the screen.
The strips are much longer than they are wide, with plasma density and width that vary only slowly with length.
Magnetic field lines provide a plausible framework for noodle-like plasma structures.

On scales smaller than those of source currents, magnetic field lines tend to lie parallel.
The Maxwell stress tensor expresses this fact in mathematical form: electric and magnetic fields exert tension along field lines, and pressure across them. 
A mnemonic holds that
field lines act like ``fuzzy rubber bands:'' they tend to contract along their length, and expand in the transverse dimensions \citep[][p. 156]{heald2012classical}.
The equilibrium configuration is thus a uniform field and parallel field lines.
Alfv\'en's theorem states that an infinitely conducting plasma cannot diffuse across field lines, because the cyclotron orbiting of charged particles ``freezes'' them to the field lines \citep[see, for example, ][Ch. 4]{davidson2001introduction}.
Because of that cyclotron orbiting, transport of charged particles along field lines is much easier than motion across them. Therefore any fluctuation of plasma density tends to be nearly constant along field lines, but to vary much more strongly across them.
Such elongated and magnetic field-aligned density fluctuations are observed via scattering in the solar wind \citep{1989JOSAA...6..977C,1990ApJ...358..685A,1997JGR...102..263G}. 
\citet{1994Natur.372..754D} observed evidence for alignment of the long axes of elliptical scattering disks with the Galactic magnetic field.
These physical considerations motivate the model shown in Figure\ \ref{fig:basic_idea}.

Of course, interstellar magnetic fields cannot be parallel everywhere. Regions where the magnetic field changes direction (more precisely, where it has nonzero curl) in a plasma give rise to thin sheet currents, as a direct consequence of Ampere's Law \citep[][Ch. 5]{davidson2001introduction}.
In these regions, breaking the field lines and reconnecting them across the sheet relieves the tension along the field lines.
This process violates Alfv\'en's Theorem; but finite resistivity of the plasma allows this.
However, in low-resistivity plasmas, such as those in interstellar and interplanetary space and some laboratory experiments, reconnection takes place orders of magnitudes faster than collisional resistivity would allow. Such collisionless reconnection likely involves the generation of strong turbulence in the current sheet, and scattering of electrons from the resulting turbulent fluctuations. This process remains the subject of intense study \citep[see, for example, ][]{gonzalez2016magnetic}.

\citet{2014MNRAS.442.3338P} suggested that reconnection sheets, corrugated so that sections are tangent to the line of sight, 
could contain sufficient electron column to produce the arcs. 
They followed a suggestion by \citet{1995ApJ...438..763G} that the interstellar turbulence responsible for scattering might lie in reconnection sheets.
\citet{1995ApJ...438..763G}, \citet{2001ApJ...554.1175M}, and \citet{2001ApJ...562..279L} suggest that the resulting fluctuations in plasma density should lie in long, thin filaments along magnetic field lines.
Periodic  plasma instabilities in the reconnection sheets might also lie parallel.
These suggestions further motivate scattering by noodles.

Turbulence involves a cascade, from large length scales where energy is injected to small scales where it is dissipated \citep{frisch1995turbulence}.
Structures of a particular size, such as vortices of a particular radius or deflections of magnetic field lines (Alfv\'en waves) of a particular wavelength, help to visualize the degrees of freedom available at a particular length scale.
Often turbulence is highly intermittent, in the sense that the turbulent cascade involves only a few of the available degrees of freedom at each scale.
In this case, filamentary density fluctuations would be rare, and most radiation would pass through the screen unscattered;
the scattering would be ``optically thin'' in the sense discussed in Sections\ \ref{sec:Intro} and\ \ref{sec:OpticalDepth}.
Indeed, observations show that scintillation arcs result from relatively rare, but relatively large, deflections.
Statistics of such scattering is that of a L\'evy flight \citep{2003ApJ...584..791B,2003PhRvL..91m1101B,2005ApJ...624..213B}.
A L\'evy flight is a random walk where the distribution of steps has a ``fat tail,'' so that outcomes are dominated by a single, large step; rather than one such as  Brownian motion, where the step lengths are drawn from a Gaussian or similar distribution and the accumulation of many small steps dominates outcomes \citep{KlafterSchlesingerZumofen1996PhT}.
L\'evy statistics are useful for modeling the stock market, blood flow, and many other such processes \citep{voit2013statistical}.

As I discuss in Section\ \ref{sec:ReferenceModel} below, 
parallel noodles lead to thin, parabolic scintillation arcs.
Parallel noodles are also in accord with interferometric observations showing highly anisotropic structures.
Misaligned noodles introduce wider structures if randomly oriented, and multiple arcs or off-arc features if not, as I discuss in Section\ \ref{sec:Extensions};
indeed a wide variety of such structure are observed\ \citep{2018MNRAS.480.4199F}.
In reconnection sheets in particular, one expects cohorts of magnetic field lines with different orientations.
If scintillation arcs arise from ``magnetic noodles,'' observations of them will allow visualization of magnetic fields in reconnection regions.

\subsection{Outline of Paper}

This paper concerns primarily the optics of noodles, as calculated by Kirchhoff diffraction of strips in a thin screen.
Figure\ \ref{fig:basic_idea} shows the geometry.
Outside of relatively long, narrow strips, localized in a thin screen, the medium is assumed to be uniform.
Physically, these strips may be filaments perpendicular to the line of sight, or sheets with normal perpendicular to the line of sight. 
This model is motivated by the work of \citet{2014MNRAS.442.3338P}, 
 \citet{2016MNRAS.458.1289L}, 
and \citet{2018MNRAS.478..983S}. 
The feature that the strips are assumed to lie parallel is further motivated by theories and observations that indicate density fluctuations in a magnetized plasma tend to be lie along aligned magnetic field lines
as discussed in the preceding Section\ \ref{sec:PhysicalNature}.
We find in this paper that parallel strips give rise to narrow scintillation arcs, with the possibility of wider or more complicated structures from canted strips, and thus reinforce this assumption.

I use the theory of Kirchhoff diffraction to find the field at the observer from an assemblage of parallel strips,
under the most general assumptions, in Section\ \ref{sec:KirchhoffIntegral}.
In Section\ \ref{sec:Definitions}, I introduce  Kirchhoff diffraction theory. 
I find the field in the absence of scattering (the ``no-screen'' case) in Section\ \ref{sec:NoScreen}, and then in Section\ \ref{sec:Strip} the change in that field from introducing a single strip. 
The integral along the strip yields a constant $\sqrt{2\pi i}\, r_\mathrm{F}$, where $r_\mathrm{F}$ is the Fresnel scale 
defined in Equation\ \ref{eq:FresnelScaleDef}.
The integral across the strip has the form of a Fourier transform, with a magnitude of order the strip width $w$,
and the geometric phase of the point where the strip lies closest to the line of sight.
The single-strip expression leads to the expression for many such strips in Section\ \ref{sec:ManyParallelStrips}.

I describe a simple reference model that yields scintillation arcs from strips in Section\ \ref{sec:OverScintillationArcsFromNoodles}.
In Section\ \ref{sec:Approximations}, I introduce a number of simplifying approximations,
and find their consequences for the Kirchhoff integral in Section\ \ref{sec:Implications}.
In Section\ \ref{sec:ReferenceModel}, I show that under these approximations, an assemblage of strips in combination with the undeflected line of sight will generically produce scintillation arcs.

In Section\ \ref{sec:TypicalParameters}, I compare results with observations to justify the approximations made earlier.
I discuss various extensions of the noodle theory in Section\ \ref{sec:Extensions}:
noodles slanted with respect to the screen plane; noodles in the screen plane but canted to the other, parallel noodles; and noodles that curve, bend, or have finite length. I also discuss wide-bandwidth, long-time, and interferometric observations.
Section\ \ref{sec:Summary} summarizes the conclusions.

\section{Kirchhoff Integral}\label{sec:KirchhoffIntegral}

\subsection{Background and Definitions}\label{sec:Definitions}

\begin{figure}
\centering
\includegraphics[width=0.45\textwidth]{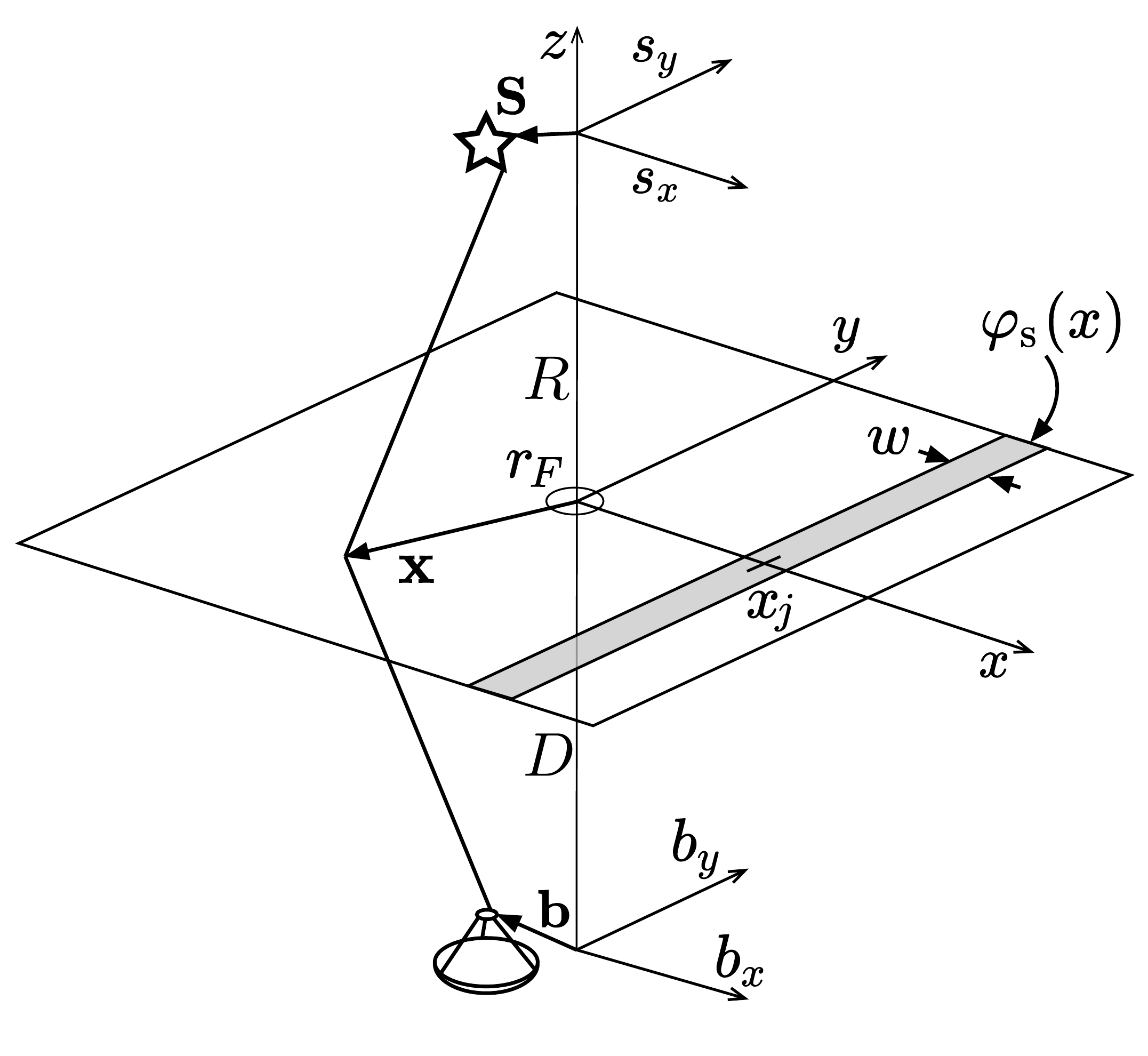}
\caption{Geometry for Kirchhoff integral, showing the path from source at $\mathbf{s}$ to observer at $\mathbf{b}$, passing through screen at a point $\mathbf{x}$. The screen introduces nonzero phase only in a thin strip of width $w$, centered at $x_j$.
\label{fig:ReferenceGeom}}
\end{figure}

Kirchhoff diffraction calculates the field at the observer as a scalar source field times three factors: a geometric phase from source to a point on the screen, an additional phase change and possible attenuation from transmission through the screen at that point, and a geometric phase from the screen point to the observer. Integration over all points on the screen (and over the source if it is extended) then yields the field at the observer \citep[see, for example,][]{goodman2017introduction}.
Kirchhoff diffraction is well suited to modeling the formation of scintillation arcs, because it accurately models electromagnetic-wave scattering at small angular deflections, in screens much thinner than the length of the line of sight, and without coupling of polarizations. The parabolic wave equation provides solution of the scalar wave equation in the more general case of small angular deflections in media extended along the line of sight\ \citep{levy2000parabolic}. For a thin screen, Kirchhoff integration solves the parabolic wave equation. 
It assumes scalar rather than vector fields; 
but for small-angle diffraction and phase changes independent of polarization, the Kirchhoff integral provides a good approximation for each component of the electric field.
Ray tracing through a thin screen, the approach used by most previous studies of scintillation arcs, is the further limit where a sum over paths of stationary phase correctly approximates the Kirchhoff integral. 
I will discuss wave optics and the ray limit, and the electron column required to reproduce the strengths of the observed scintillation arcs, in Paper II.

For a field at the source $\psi_\mathrm{src} (\mathbf{s})$, the scattered field at the observer $\psi_\mathrm{obs} (\mathbf{b})$ is
\citep[see, for example][]{2015ApJ...805..180J}:
\begin{align}
\label{eq:KirchhoffIntegral}
\psi_\mathrm{obs} (\mathbf{b}) &= \frac{-i}{2\pi r_\mathrm{F}^2} \int_{\rm screen} d^2x\; e^{i\left[ \left( \frac{k}{2 D}\right) |\mathbf{b} - \mathbf{x} |^2 + \varphi(\mathbf{x}) \right]} 
\int_{\rm source} d^2s  \; e^{i\left( \frac{k}{2R}\right) |\mathbf{x} - \mathbf{s} |^2} \psi_\mathrm{src} (\mathbf{s})
.
\end{align}
Figure\ \ref{fig:ReferenceGeom} shows the geometry.
The wavenumber is $k=2\pi \nu/c$, where $\nu$ is the observing frequency. 
The optical axis is the $z$-axis.
The distance from observer to screen is $D$, and that from screen to pulsar is $R$.
The transverse coordinates are $\mathbf{b}$ in the observer plane, $\mathbf{x}$ in the screen plane, and $\mathbf{s}$ in the source plane.
The scattering screen introduces a phase change by $\varphi(\mathbf{x})$.
In analogy to the magnification of the scattering screen viewed as a lens, I define:
\begin{align}
M &= \frac{D}{R} 
.
\end{align}
The Fresnel scale is the lateral separation at the scattering screen that produces an additional geometric path length of one-half radian of phase.
That scale is:
\begin{align}
\label{eq:FresnelScaleDef}
r_\mathrm{F} &=\sqrt{ \frac{DR}{(D+R)} \frac{1}{k} } 
= \sqrt{\frac{1}{(1+M)}\frac{D}{k}} 
.
\end{align}
I suppose that the source is pointlike, so that the integral of $\psi_\mathrm{src} $ over over the source plane collapses to a single value, at the point $\mathbf{s}$.
The Kirchhoff integral then takes the form:
\begin{align}
\label{eq:PsiFresnelPointSource}
\psi_\mathrm{obs} (\mathbf{b}) &= \frac{-i}{2 \pi r_\mathrm{F}^2} \, e^{ \frac{i}{2 r_\mathrm{F}^2 } \frac{1}{(1+M)}\left( |\mathbf{b}|^2 + M |\mathbf{s}|^2 \right) }
\int_{\rm screen} d^2x\; e^{\frac{i}{2 r_\mathrm{F}^2 } \left[ -2 \frac{ ( \mathbf{b}+ M \mathbf{s})}{(1+M) } \cdot \mathbf{x} + |\mathbf{x}|^2 + \varphi(\mathbf{x}) \right]} \psi_{{\rm src}}
,
\end{align}
where I display the quadratic geometric phases at source and observer outside the integral.
These quadratic phases express the fact that the observing plane and the source plane are flat, rather than curved like spherical wavefronts centered on the screen at $\mathbf{x}=0$. 
The position of the undeflected line of sight in the screen plane, sometimes known as the ``scaled baseline,'' is:
\begin{align}
\mathbf{x}_\mathrm{los} &\equiv \mathbf{b}_1 = \frac{\mathbf{b} + M \mathbf{S}}{(1+M)}
.
\end{align} 
All dependence of the integral in Equation\ \ref{eq:PsiFresnelPointSource} on source and observer position is through $\mathbf{b}_1$.

The Fresnel phase is:
\begin{align}
\phi_\mathbf{F} &=\frac{|\mathbf{x}|^2}{2 r_\mathrm{F}^2}
\end{align}
This is the phase introduced by geometric path length, for a path that passes from a source on the optical axis to an observer on the optical axis, but through a point on the screen at $\mathbf{x}$.  A ``Fresnel zone'' is the annulus between $\phi_\mathrm{F}=(N+\frac{1}{2})\pi$ and $\phi_\mathrm{F}=(N+\frac{3}{2})\pi$, where $N$ is an integer (except for the first Fresnel zone, which extends from $\phi_\mathrm{F}=0$ to $\phi_\mathrm{F}=\frac{1}{2}\pi$). Thus, across a pair of adjacent Fresnel zones, the Fresnel phase changes by $2\pi$.
The width of a pair of Fresnel zones is:
\begin{align}
\label{eq:WidthOfAPairOfFresnelZones}
2\pi \left(\left.  \frac{\partial \phi_\mathrm{F}}{\partial x} \right|_{x_j} \right)^{-1} = \frac{ 2\pi r_\mathrm{F}^2}{x_j}
\end{align}
where $x_j$ is the distance from the optical axis on the screen to the pair.

\subsection{No Screen}\label{sec:NoScreen}

For reference and to introduce integrals, I consider the case of zero screen phase: $\varphi_\mathrm{s}=0$.
This is the case of no screen at all.
For this calculation only, without loss of generality, I take $\hat{\mathbf{x}}$ parallel to $ \mathbf{b}+  M \mathbf{s} $,
so that the $y$-component of the scaled baseline vanished: $b_{1y}=b_y+ M s_y =0$.
Equation\ \ref{eq:PsiFresnelPointSource} then takes the form:
\begin{align}
\psi_\mathrm{NS} (b \hat x) &= \frac{-i}{2 \pi r_\mathrm{F}^2} \, e^{ \frac{i}{2 r_\mathrm{F}^2 } \frac{1}{(1+M)}\left( |\mathbf{b}|^2 + M |\mathbf{s}|^2 \right) } \int_{-\infty}^\infty dx \; e^{ \frac{i}{2 r_\mathrm{F}^2}  \left( - 2 b_{1x}  x+x^2 \right) } \int_{-\infty}^\infty dy \; e^{ \frac{i}{2 r_\mathrm{F}^2} y^2 } \psi_{{\rm src}}
\end{align}
where $b_\mathrm{1x}=(b_\mathrm{x}+M s_\mathrm{x} )/(1+M)$.
The integral over $y$ is that of a Gaussian function with imaginary variance.
One can use analytic continuation to extend the integral of a Gaussian function with positive real part for the variance to this case.
Alternatively, one can invoke the stationary-phase approximation; 
indeed, this integral is the archetype for integration using the stationary-phase approximation \cite[see][]{bender2013advanced}.
The integral over $y$ thus becomes:
\begin{align}
\int_{-\infty}^\infty dy \; e^{ \frac{i}{2 r_\mathrm{F}^2} y^2 } &= \sqrt{2\pi} \, r_\mathrm{F} e^{i\frac{\pi}{4}}
\end{align}
I complete the square to convert the integral over $x$ to the same form, times a phase factor:
\begin{align}
\label{eq:NoScreenIntegralOverX}
\int_{-\infty}^\infty dx \; e^{ \frac{i}{2 r_\mathrm{F}^2}  \left( -\frac{2}{(1+M)} ( b_x +M s_x ) x+x^2 \right) }
=  \sqrt{2\pi}\, r_\mathrm{F} e^{-\frac{i}{2 r_\mathrm{F}^2}b_\mathrm{1x}^2+ i\frac{\pi}{4}}
.
\end{align}
I then find for the field at the observer at $\mathbf{b}$ :
\begin{align}
\label{eq:psiNoScreen}
\psi_\mathrm{NS} 
&= e^{\frac{i}{2 r_\mathrm{F}^2} \frac{M| \mathbf{b}-\mathbf{s} |^2}{(1+M)^2} } \, \psi_\mathrm{src} = e^{i \frac{k}{D+R} | \mathbf{b}-\mathbf{s} |^2} \, \psi_\mathrm{src}
\end{align}
where I make use of the condition that $b_y+ M s_y =0$ to obtain a coordinate-independent form,
and eliminate $M$ and $r_\mathrm{F}$ in favor of $k$, $D$, and $R$.
Note that in the absence of a screen,
the magnitude of the observed field is $|\psi_\mathrm{src}|$, and the phase of the observed field depends only on the lateral separation of source and observer $|\mathbf{b}-\mathbf{s}|$, as expected.
For later convenience, I define this phase as:
\begin{align}
\label{eq:phiNSdef}
\phi_\mathrm{NS} &\equiv {\frac{1}{2 r_\mathrm{F}^2} \frac{M| \mathbf{b}-\mathbf{s} |^2}{(1+M)^2} }
\end{align}
The field is independent of the position of the screen, as required; and its magnitude is independent of the distance $D+R$, for the Kirchhoff integral as normalized in Equation\ \ref{eq:KirchhoffIntegral}.
The intensity at the observer at $\mathbf{b}$ is simply:
\begin{align}
I_{\rm NS}  &= \psi_\mathrm{NS}  \psi_\mathrm{NS}^* = |\psi_{{\rm src}}|^2
\end{align}
also as expected.

\subsection{Strips}\label{sec:Strip}

I now suppose that the entire screen plane introduces zero phase change, except within a narrow strip of width $w$, where the phase takes the form $\varphi_\mathrm{s}(x)$.
Without loss of generality I suppose that the strip lies perpendicular to the $x$-axis with its midline at an offset $x_j$ from the optical axis.
I assume that $\varphi_\mathrm{s}(x)$ depends only on the coordinate perpendicular to the width of the strip, $x$.
The phase is uniform along the strip, in $y$.
Figure\ \ref{fig:ReferenceGeom} shows the geometry.
I argued that the ``freezing'' of plasma fluctuations to magnetic field lines, and easy diffusion along them, suggests this geometry in Section\ \ref{sec:PhysicalNature}.

I divide the integral over the screen plane into 3 parts: 
the strip with phase $\varphi_\mathrm{s}(x)$, contributing $\psi_\mathrm{s} $ to the field at the observer;
the integral over the entire screen with zero phase, with value $\psi_\mathrm{NS}$;
and the contribution of the strip with zero phase, $\psi_\mathrm{s0} $, to be subtracted from the other two:
\begin{align}
\label{eq:TriDissection}
\psi_\mathrm{obs} &= \psi_\mathrm{NS} + \psi_\mathrm{s} - \psi_\mathrm{s0}
\end{align}

\subsubsection{Single Strip}

The contribution of the strip $\psi_\mathrm{s}$ is given by integration over its portion of the domain of $x$ in Equation\ \ref{eq:PsiFresnelPointSource}:
\begin{align}
\label{eq:DiffractiveStrip}
\psi_{\rm s} &= \frac{-i}{2 \pi r_\mathrm{F}^2} 
\, e^{ \frac{i}{2 r_\mathrm{F}^2 } \frac{1}{(1+M)}\left( |\mathbf{b}|^2 + M |\mathbf{s}|^2 \right) }
\int_{x_j-w/2}^{x_j+w/2} dx \;  e^{ \frac{i }{2 r_\mathrm{F}^2}  \left( -2 b_{1x} x+x^2 \right) } 
\int_{-\infty}^\infty dy \; e^{\frac{i}{2 r_\mathrm{F}2} \left( -2 b_{1y} y + y^2 \right) } e^{i\varphi_\mathrm{s} (x) } \psi_\mathrm{src}  
\end{align}
The integral over $y$ is identical to the integral over $x$ in the case of no screen, Equation\ \ref{eq:NoScreenIntegralOverX}. Thus, it contributes a factor of magnitude $\sqrt{2 \pi}\, r_\mathrm{F}$ and of phase $( b_\mathrm{1y}^2/(2 r_\mathrm{F}^2) + \pi/4 )$, so that:
\begin{align}
\psi_{\rm s} 
&=  \frac{1 }{\sqrt{2 \pi}\, r_\mathrm{F}} 
\, e^{ \frac{i}{2 r_\mathrm{F}^2 } \left(  \frac{1}{(1+M)}\left( |\mathbf{b}|^2 + M |\mathbf{s}|^2 \right) - b_\mathrm{1y}^2  \right)- i \frac{\pi}{4}}
\int_{x_j-w/2}^{x_j+w/2} dx \;  e^{ \frac{i }{2 r_\mathrm{F}^2}  \left( -2 b_\mathrm{1x} x+x^2 \right) } e^{i\varphi_\mathrm{s} (x) }
\psi_\mathrm{src}  
.
\end{align}
I use the substitution $u= x - x_j$ to find for the contribution of the strip to the field:
\begin{align}
\label{eq:DiffractiveStripSimplified}
\psi_\mathrm{s} 
&=\frac{1}{\sqrt{2\pi} r_\mathrm{F}}  
\, e^{ \frac{i}{2 r_\mathrm{F}^2 } \left(  \frac{1}{(1+M)}\left( |\mathbf{b}|^2 + M |\mathbf{s}|^2 \right) - b_\mathrm{1y}^2  -2  b_\mathrm{1x} x_j + x_j^2 \right)- i \frac{\pi}{4}}
\int_{-w/2}^ {+w/2} du\, e^{\frac {i}{2r_\mathrm{F}^2} (2(- b_\mathrm{1x} + x_j)  u +u^2)} e^{i\varphi_\mathrm{s} (u)} \psi_\mathrm{src}  \\
&=\frac{1}{\sqrt{2\pi} r_\mathrm{F}}  
\, e^{ i \phi_{\mathrm{g}j} - i \frac{\pi}{4}}
\int_{-w/2}^ {+w/2} du\, e^{\frac {i}{2r_\mathrm{F}^2} (2(- b_\mathrm{1x} + x_j)  u +u^2)} e^{i\varphi_\mathrm{s} (u)} \psi_\mathrm{src}  
\end{align}
where $\phi_{\mathrm{g}j}$ is the geometric phase at $x_j$, the center of the strip in $x$, where the strip passes closest to the line of sight in $y$:
\begin{align}
\label{eq:FullPhiGeomDef}
\phi_{\mathrm{g}j} &= \frac{1}{2 r_\mathrm{F}^2 } \left(  \frac{1}{(1+M)}\left( |\mathbf{b}|^2 + M |\mathbf{s}|^2 \right) - b_\mathrm{1y}^2  -2  b_\mathrm{1x} x_j + x_j^2 \right)
 =  \frac{1}{2 r_\mathrm{F}^2 } \left( \frac{ M (b_y-s_y)^2}{(1+M)^2} +\frac{ (b_x - x_j)^2+ M (s_x-x_j)^2}{(1+M)} \right)
\end{align}
The subscript $j$ indicates that $\phi_{\mathrm{g}j}$ depends on $x_j$.
Note that $\phi_{\mathrm{g}j}$ contains three sorts of terms:
those that are quadratic in the offset $x_j/r_\mathrm{F}$; those that are quadratic in, or a product of, the displacements $b/r_\mathrm{F}$ and $s/r_\mathrm{F}$; and those that are linear in both $x_j/r_\mathrm{F}$ and $b/r_\mathrm{F}$ or $s/r_\mathrm{F}$.
Variation of the geometric phase with frequency through $r_\mathrm{F}$, and with time through $b$ and $s$, gives rise to scintillation arcs,
as we demonstrate in Section\ \ref{sec:ReferenceModel} below.

The field at the observer is then given by Equation\ \ref{eq:TriDissection}: it is the sum of the contribution of the screen without a strip, $\psi_\mathrm{NS}$; 
the contribution of the strip, $\psi_\mathrm{s}$;
and the negative of the contribution of the strip with zero screen phase, $\psi_\mathrm{s0}$:
\begin{align}
\psi_\mathrm{obs} &= \psi_\mathrm{NS} + \psi_\mathrm{s} - \psi_\mathrm{s0}
\nonumber \\
\label{eq:NetPsiFromStripToBeFourierTransformed}
&
=e^{\frac{i}{2 r_F^2}\frac{M | \mathbf b - \mathbf s |^2}{(1+M)^2} }\psi_\mathrm{src}   + \frac{1}{\sqrt{2\pi} r_\mathrm{F}} e^{ i \phi_{\mathrm{g}j} - i \frac{\pi}{4}}  \left( \int_{-w/2}^ {+w/2} du\, e^{\frac {i}{2r_\mathrm{F}^2} (2 (- b_\mathrm{1x} +x_j) u +u^2)} \left[ e^{i\varphi_\mathrm{s} (u)}  - 1 \right] \right) \psi_\mathrm{src} 
.
\end{align}
This is a fundamental result of this paper.

\subsubsection{Fourier transform}\label{sec:FourierTransform}

One can place Equation\ \ref{eq:NetPsiFromStripToBeFourierTransformed} into a somewhat more intuitive form by associating the phase quadratic in $u$ with the screen phase, and expressing the limits of the integral as a boxcar function:
\begin{align}
\label{eq:NetPsiFromStrip}
\psi_\mathrm{obs} 
&=\psi_\mathrm{NS}   +  \frac{1}{\sqrt{2\pi} r_\mathrm{F}}  e^{ i \phi_{\mathrm{g}j} - i \frac{\pi}{4}} \left( \int_{-\infty}^ {+\infty} du\cdot e^{\frac {i}{r_\mathrm{F}^2} (- b_\mathrm{1x}    +x_j) u} 
\cdot B_w (u) 
\cdot \left[  e^{i\varphi_\mathrm{s} (u) + \frac {i}{2r_\mathrm{F}^2}u^2 }  - e^{ \frac {i}{2r_\mathrm{F}^2}u^2 }  \right] \right) \psi_\mathrm{src} 
\end{align}
where the boxcar function is:
\begin{align}
B_w(u) &= 
\begin{cases}
1 & -w/2 < u < w/2 \\
0 & \mathrm{otherwise .}
\end{cases}
\end{align}
The term in braces $(...)$ in Equation\ \ref{eq:NetPsiFromStrip} has the form of a Fourier transform.
The integral Fourier transforms from the domain of the variable 
\begin{align}
u = x-x_j
\end{align}
to the variable
\begin{align}
\label{eq:q_jDefn}
q_j&\equiv \frac{1}{r_\mathrm{F}^2}(- b_\mathrm{1x} + x_j) =\frac{1}{r_\mathrm{F}^2} \left(-\frac{( b_x+M s_x )}{(1+M) } +x_j \right)
\end{align}
Thus, the Fourier transform converts from the screen plane to the observer plane, as is the case in Fourier optics generally \citep{goodman2017introduction}.
The subscript on the variable $q_j$ indicates that, like the geometric phase $\phi_{\mathrm{g}j}$, it depends on $x_j$.
The function to be transformed is the product of the boxcar function and the term in square brackets $[...]$ in Equation\ \ref{eq:NetPsiFromStrip}.
I define that term to be the function $g(u)$:
\begin{align}
\label{eq:gDef}
g(u) &\equiv e^{i\varphi_\mathrm{s} (u) + \frac {i}{2r_\mathrm{F}^2}u^2 }  - e^{ \frac {i}{2r_\mathrm{F}^2}u^2 }  
\end{align}
The Fourier transform of the boxcar function is:
\begin{align}
\tilde B_w(q) &= w \, \mathrm{sinc} \left( \frac{ w q }{2} \right)
,
\end{align}
where the tilde $\tilde\ $ denotes the Fourier transform,
and the sinc function is defined as $\mathrm{sinc} (t ) \equiv \sin(t)/t$.
The Fourier transform of $g(u)$ is $\tilde g (q)$.
The Fourier transform of the product $B_w(u) \cdot g(u)$ is the convolution $\tilde B_w(q) \Conv \tilde g(q)$.
The convolution must be evaluated at $q_j$, as defined in Equation\ \ref{eq:q_jDefn}.
I thus write the observed field in the form:
\begin{align}
\label{eq:psi_obs_GeneralForm}
\psi_\mathrm{obs} &= \psi_\mathrm{NS}   + \frac{w}{\sqrt{2\pi} r_\mathrm{F}}  e^{ i \phi_{\mathrm{g}j} - i \frac{\pi}{4}}  \left[ \mathrm{sinc} \left( \frac{ w q }{2} \right) \Conv \tilde g(q ) \right]_{q_j} \psi_\mathrm{src} \\
\label{eq:PhasorForOneStrip}
&\equiv  \psi_\mathrm{NS}   + \Gamma_j e^{i  \phi_{\mathrm{g}j}}  \psi_\mathrm{src} 
,
\end{align}
where I have defined the structure factor
\begin{align}
\label{eq:GammaDef}
\Gamma_j &\equiv \frac{w}{\sqrt{2\pi} r_\mathrm{F}}  e^{ - i \frac{\pi}{4}}  \left[ \mathrm{sinc} \left( \frac{ w q }{2} \right) \Conv \tilde g(q ) \right]_{q_j} 
.
\end{align}
All dependence of the scattered field on the structure of the strip is contained in $\Gamma_j$. 
The sinc function has its first zeros at $w q_j/2=\pm \pi$, so its characteristic width in the $q_j$-domain is 
\begin{align}
\label{eq:WqDef}
W_q = 4\pi/w
.
\end{align}
Because of the convolution, the structure factor $\Gamma_j$ is smooth on scales of $W_q$ or smaller, in the domain of $q_j$. 
I will explore the variation of $\Gamma_j$ with frequency and time in Section\ \ref{sec:InternalStructureUncertaintyPrinciple} below.

\subsubsection{Many Parallel Strips}\label{sec:ManyParallelStrips}

To generalize to the field in the case where the screen contains multiple, parallel strips, I simply sum the second term of Equation\ \ref{eq:PhasorForOneStrip} over $j$:
\begin{align}
\label{eq:MultiStripFieldGeneral}
\psi_\mathrm{obs} &=  \psi_\mathrm{NS}   + \sum_j \left( \Gamma_j e^{i  \phi_{\mathrm{g}j}}  \right) \psi_\mathrm{src}
\end{align}
Those strips may have different widths $w_j$, as well as different forms for their internal phases $\phi_s(x)$, leading to different coefficients $\Gamma_j$.  Moreover, at this stage of the calculation, before approximations, $\Gamma_j$ and $\phi_{\mathrm{g}j}$ depend on frequency and on the positions of source and observer; because source and observer are in motion, they depend on time.
The resulting intensity at the observer is the square modulus of the field:
\begin{align}
\label{eq:MultiStripIntensityGeneral}
I_\mathrm{obs} &= \psi_\mathrm{obs} \psi_\mathrm{obs}^* = \left| \psi_\mathrm{src}\right|^2 \left( \left[ 1 + \sum_j \left| \Gamma_j \right|^2 \right]
+ 2 \sum_j \mathrm{Re}\left[ \Gamma_j e^{i (\phi_{\mathrm{g}j}-\phi_\mathrm{NS})} \right]
+ 2 \sum_{k<j} \mathrm{Re}\left[ \Gamma_j \Gamma_k^* e^{i (\phi_{\mathrm{g}j}-\phi_{\mathrm{g}k}) }  \right]  \right)
\end{align}
where Equation\ \ref{eq:phiNSdef} defines $\phi_\mathrm{NS}$.
This is an important result of this paper. I show below that the first term in rectangular brackets $[...]$ is responsible for the central maximum  ($a$, in Figure\ \ref{fig:reference_arcs}), the second term is responsible for the primary arcs ($b$, $c$), and the third term is responsible for the secondary arclets ($d$, $e$).
Note that the first term is of order the intensity of the unscattered source, the second is smaller by a factor of order $\Gamma$, and the third by $\Gamma^2$;
and that $\Gamma$ is of order $w/x_j$, as Equation\ \ref{eq:GammaDef} states.
Thus, if $w\ll r_\mathrm{F}$, as we argue below, the second term is much smaller than the first, and the third smaller than the second. The terms also have distinct dependences on the geometric phase $\phi_{\mathrm{g}j}$.

In practice, the ``undeflected line of sight'' may not exist; all paths may be deflected to some degree, although most source intensity arrives via paths with very small deflection.  Indeed, Levy flights have this character: a few outcomes involve extreme events, but most outcomes experience small changes. 
In this case, the first and second terms of Equation\ \ref{eq:MultiStripIntensityGeneral} are absent.
However, as we argue in Section\ \ref{sec:OpticalDepth} below, the number, or the strength $\left| \Gamma_j \right|$, of paths near zero deflection $x_j=0$ can nevertheless make the largest contribution to the sum in the third term, by far. Consequently, the secondary arclets near the origin in the $(\tau, f)$ plane are by far the strongest. This dominance of less-scattered paths also ensures that the central part of each secondary arclet is bright enough to delineate the missing primary arc. Thus, paths with nearly zero deflection can play the role of the undeflected line of sight.

The interferometric visibility takes the form of a similar expression, but with different values for $\phi_{\mathrm{g}j}$ at the two ends of the baseline, so that 
the second and third terms become complex. Specifically, the third term becomes the sum of two complex exponentials, with phases of the difference of $\phi_{\mathrm{g}j}$ at one end of the baseline and $\phi_{\mathrm{g}k}$ at the other, and the reverse.
In principle, $\Gamma_j$ for a single value of $j$ can be different at the two ends of the baseline, although this is likely to be important only for observations spanning wide bandwidths, long times, or the longest baselines.
We discuss such observations further in Sections\ \ref{sec:WideBandLongTime} and \ \ref{sec:Interferometric} below.

\section{Scintillation Arcs from the Noodle Model}\label{sec:OverScintillationArcsFromNoodles}

In this section I investigate the case of greatest interest and applicability to the problem of scintillation arcs:
where the strip is narrow relative to variations of geometric phase,
far from the undeflected path,
and introduces a small phase change relative to the geometric phase.
I suppose in this Section \ref{sec:OverScintillationArcsFromNoodles} only that the typical observed frequency range is small compared with the observing frequency, and that the displacement of the line of sight during an observation is small compared with the offset $x_j$ of the strip from the optical axis.
From these I obtain simple forms for the observed field and intensity, that exhibit the properties of scintillation arcs.
I discuss these assumptions formally in Section\ \ref{sec:Approximations},
their implications in Section\ \ref{sec:Implications},
and construct scintillation arcs in Section\ \ref{sec:ReferenceModel}.

\subsection{Approximations}\label{sec:Approximations}

This section introduces assumptions that allow easy construction of scintillation arcs.
I show that these assumptions are justified in the case of observed scintillation arcs in Section\ \ref{sec:TypicalParameters},
and discuss the observational consequences of dropping the approximations in Sections\ \ref{sec:WideBandLongTime} and\ \ref{sec:Interferometric}.

\subsubsection{Strip Offset $x_j$}\label{sec:Approx:StripOffset}

In most interesting cases, the strip lies far outside the first Fresnel zone:
\begin{align}
\label{eq:Approx:StripOffset}
x_j \gg r_\mathrm{F}
\end{align}
As I discuss in Section\ \ref{sec:TypicalParameters},
analyses of observations cover $x_j$ in the range between $20 r_\mathrm{F}$ and $1000 r_\mathrm{F}$.
Hence I assume that $x_j \gg r_\mathrm{F}$; or, more specifically, that $x_j > 20 r_\mathrm{F}$

\subsubsection{Number of Strips}

I assume that the number of strongly scattering strips is small.
By ``strongly scattering'' I mean those that contribute with $x_j >20\ r_\mathrm{F}$, as discussed in the preceding section.
The characteristic geometric path length introduced by such a strip is 
\begin{align}
c \tau_j = \frac{c}{2\pi} \frac{\partial}{\partial \nu} \phi_\mathrm{F} =
\frac{x_j^2}{ 4\pi r_F^2}\frac{c}{\nu}
\end{align}
The assumption for strip offset, that $x_j \gg r_\mathrm{F}$, thus implies that $c \tau_j$ is much greater than a wavelength $\lambda=c/\nu$. This assumption produces strong cancellation or reinforcement of the scattered paths, relative to one another and to the undeflected line of sight.
This assumption is slightly different from the traditional assumption of strong scattering: that every path suffers a 
change in path length much larger than a wavelength, relative to an undeflected path. Here I assume that strongly scattering strips are relatively uncommon in the screen plane, so that most of the power remains 
in the undeflected path, or in common but much less-strongly scattering structures. However, the relatively few strongly-deflected paths are longer by many wavelengths.

\subsubsection{Strip Width $w$}\label{sec:Approx:StripWidth}

I consider cases where the strip is no wider than a few pairs of Fresnel zones.
Thus, I demand that:
\begin{align}
w \lesssim  \frac{ 2\pi r_\mathrm{F}^2}{x_j} 
\end{align}
where Equation\ \ref{eq:WidthOfAPairOfFresnelZones} gives the width of a pair of Fresnel zones.
We use the assumption of large screen offset in Section\ \ref{sec:Approx:StripOffset} to rewrite this in the form:
\begin{align}
w \lesssim \frac{2\pi}{20} r_\mathrm{F}
.
\end{align}
Wider strips can easily be modeled as superpositions of narrower strips.
However, the contributions of wide strips will tend to cancel out over the strip, unless the screen phase $\varphi_\mathrm{s}$ tracks the Fresnel phase $\phi_\mathrm{F}$;
I discuss this effect, in the context of specific structures for the strips, in Paper II. 

\subsubsection{Strip Phase $\varphi_\mathrm{s}$}\label{sec:Approx:StripPhase}

We suppose that the phase of the strip is small compared with the geometric phase:
$\varphi_\mathrm{s} \ll \phi_\mathrm{g}$.
The variation of the geometric phase with frequency over the band gives the delay that defines the scintillation arcs, as we discuss in Section\ \ref{sec:ReferenceModel} below.
However, strip phase also depends on frequency, through the dispersive refraction of plasma: 
\begin{align}
\varphi_\mathrm{s} = \frac{c r_0}{\nu} \int N_e\, dz
,
\end{align}
where $r_0=2.8\times 10^{-13}\ \mathrm{cm}$ is the classical radius of the electron, $N_e$ is the number density of electrons, and the integral is through the screen. 
If comparable to or greater than the geometric phase, strip dispersion will alter the shapes of the scintillation arcs. 
Thus, we demand:
\begin{align}
\left| \left[ \frac{\partial}{\partial \nu} \varphi_\mathrm{s} \right]_{\nu_0} \right| &\ll \left| \left[ \frac{\partial}{\partial \nu} \phi_g \right]_{\nu_0} \right| 
.
\end{align}
Evaluation leads to the condition:
\begin{align}
\left|  \varphi_\mathrm{s} \right|  &\ll \left|  \phi_g \right| 
.
\end{align}
Thus, if the phase change for the strip is much less than the geometric phase, dispersion of the strip can be ignored.
All present models make this assumption.

\subsubsection{Bandwidth $B$}\label{sec:Approx:Bandwidth}

An observer of strong scintillation sees large changes in intensity with changes in observing frequency.
Observations cover a range of frequencies $\nu$ within a passband of bandwidth $B$, from $\nu_0 - B/2$ to $\nu_0 + B/2$.
Here the frequency at the center of the observing band is $\nu_0$, and the offset of a particular frequency from that center is $\Delta\nu\equiv \nu-\nu_0$.  The observing frequency affects only the Fresnel scale, with value at frequency offset $\Delta\nu$ of:
\begin{align}
r_\mathrm{F}^2(\Delta\nu) &= r_{\mathrm{F}0}^2\left(1 + \frac{\Delta\nu}{\nu_0}\right)
\end{align}
where $r_{\mathrm{F}0}$ is the Fresnel scale at $\nu_0$.

In Sections\ \ref{sec:Implications} and\ \ref{sec:ReferenceModel}, I suppose that $\Delta\nu/\nu_0 \ll 1$. I will compare this with observing parameters for some current observations 
in Section\ \ref{sec:TypicalParameters} below, and then discuss predictions for wide-bandwidth observations with $\Delta\nu/\nu_0\approx 1$ in Section\ \ref{sec:WideBandLongTime}.

\subsubsection{Time Range $T$}\label{sec:Approx:TimeRange}

Over the time span of an observation, the observer at $\mathbf{b}$ and source at $\mathbf{s}$ move at constant velocities,
moving the undeflected line of sight relative to the scatterers.
The most important change in position is through the $x$-component of $\frac{d}{dt}\mathbf{b}_{1}$, perpendicular to the strip:
\begin{align}
V_x &  \equiv \frac{d}{dt} b_{1x} = \frac{1}{(1+M) } \left(  \frac{d b_x}{dt} + M \frac{d s_x}{dt} \right) ,
\end{align}
and so,
\begin{align}
b_{1x} &= V_x\, t
,
\end{align}
where I set coordinates so that the source and observer are on the optical axis, at $\mathbf{s}=0$ and $\mathbf{b}=0$, at $t=0$.
We assume that, over the course of an observation, the source and observer move by no more than a few Fresnel scales. Thus, during a single observation,
\begin{align}
\label{eq:Approx:TimeSpan}
|V_x t| \le V_x T/2 \lesssim  r_\mathrm{F}
.
\end{align}
Note that this implies that $|V_x t| \ll x_j$, by Equation\ \ref{eq:Approx:StripOffset}.
We make these assumptions in Sections\ \ref{sec:Implications} and\ \ref{sec:ReferenceModel},
and present theoretical results for observations that violate them in Section\ \ref{sec:WideBandLongTime}.
As we discuss in Section\ \ref{sec:TypicalParameters}, the assumptions hold for observations of B0834+06 by \citeauthor{2010ApJ...708..232B}
and for observations of pulsar J0427-4715 at $\nu_0=732\ MHz$ by \citeauthor{Reardon2018PhDT}.
For observations of pulsar J0427-4715 at $\nu_0=1400\ MHz$ by \citeauthor{Reardon2018PhDT}, $V_x T/2 \approx 10 r_\mathrm{F}$.
These present an interesting test case for a long-time observation, as discussed further in Section\ \ref{sec:WideBandLongTime}.

\subsection{Consequences of Approximations}\label{sec:Implications}

\subsubsection{Relative Intensities, Optical Depth to Scattering, and Selection Effect}\label{sec:OpticalDepth}

Recall that the intensity of the scintillation pattern consists of three terms
involving different powers of $|\Gamma | \approx w/r_\mathrm{F}$, as Equation\ \ref{eq:MultiStripIntensityGeneral} shows.
We assume in Section\ \ref{sec:Approx:StripWidth} that the strip is narrow, $w\ll r_F$, so the sum in the second term of Equation\ \ref{eq:MultiStripIntensityGeneral} is smaller than the first term by a factor of $w/r_\mathrm{F}$;
and the third term is smaller than the second by the same factor.
These three terms correspond to the undeflected path, the primary arc, and the secondary arclets, as shown in Figure\ \ref{fig:reference_arcs}. For the most striking scintillation arcs, it appears that the undeflected path does indeed include most of the intensity,
while the primary arcs are weaker, and the secondary arclets are weaker yet, as the figure suggests
\citep{2010ApJ...708..232B,2011ApJ...733...52G}.

In practice, the undeflected path may not exist.
In this case, the  first and second terms in Equation\ \ref{eq:MultiStripIntensityGeneral} are absent.
A selection effect will nevertheless 
form a numerous powerful set of paths with small deflections, that plays the role of the undeflected line of sight.
The effect arises from the fact that a strip scatters most effectively if its width is less than or about that of a pair of Fresnel zones,
$w\lesssim 2\pi r_\mathrm{F}^2$.
The effect is particularly pronounced for small electron columns $|\varphi_\mathrm{s}| \lesssim 2$ and smooth variation of $\varphi_\mathrm{s}$ within the strip;
for wider strips, 
parts of the strip with nearly the same $\varphi_\mathrm{s}$ but opposite Fresnel phases will cancel in the integration over $u=x-x_j$.
(I discuss this quantitatively, in the context of a specific model for $\varphi_\mathrm{s}$, in Paper II.)
If the strip is narrower than this limit, it will make a contribution to the scattered field of order $|\Gamma_j|\approx {w}/\sqrt{2\pi} r_\mathrm{F}$.
The scattered field from the strip is thus:
\begin{align}
\frac{|\psi_\mathrm{s} |}{ | \psi_\mathrm{src} | } \approx \begin{cases}
\frac{w}{\sqrt{2\pi} r_\mathrm{F}}& \mathrm{if\ } |x_j| < \frac{2\pi r_\mathrm{F}^2}{w} \\
0 &\mathrm{otherwise.}
\end{cases}
\end{align}
where I have used Equations\ \ref{eq:GammaDef} and\ \ref{eq:WidthOfAPairOfFresnelZones}.
Therefore, a strip of a given width $w$ will appear only over a range of offsets $x_j$ inversely proportional to its width.
Thus, narrow strips can appear rather far away, but will be faint because they are narrow;
wider strips can appear only when they are closer to the undeflected line of sight, and will be brighter because they are wider.
This ensures that a population of wide strips can play the role of the (theoretical) undeflected line of sight.
It will also ensure that the ``cloud'' of scattered paths responsible for the arc remains close to the pulsar, as it moves across the sky.
In Paper II, I show that a distribution of widths and plasma columns can reproduce the distribution of arcs and their intensity as a function of delay and rate; or equivalently as a function of position on the sky $x_j/D$. 

\subsubsection{Geometric Phase $\phi_{\mathrm{g}j}$}

The geometric phase $\phi_{\mathrm{g}j}$ is large and can vary rapidly with frequency and time.
We assume here and in Section\ \ref{sec:ReferenceModel} that observing bandwidth is small compared with observing frequency, $\Delta\nu\ll \nu_0$,
and that displacements are small compared with the Fresnel scale, $V_x T/2 \ll r_\mathrm{F}$, as discussed in Sections\ \ref{sec:Approx:Bandwidth} and\ \ref{sec:Approx:TimeRange}. 
Therefore, 
in the definition of $\phi_{\mathrm{g}j}$, Equation\ \ref{eq:FullPhiGeomDef}, we can neglect contributions that are quadratic in displacements: $b^2/r_\mathrm{F}^2$, $s^2/r_\mathrm{F}^2$, and so on;
and terms that are proportional to frequency offset times a displacement: $(\Delta\nu/\nu_0) b/r_\mathrm{F}$, $(\Delta\nu/\nu_0) s/r_\mathrm{F}$, and so on.
On the other hand, because we assume that the strip lies far outside the first Fresnel zone, $x_j \gg r_\mathrm{F}$,
contributions that are quadratic in $x_j/r_\mathrm{F}$ will be very large, and we must include the product with $\Delta\nu/\nu_0$.
We apply these assumptions to find:
\begin{align}
\label{eq:ApproxPhiGeomDefFreqVariation}
\phi_{\mathrm{g}j} &\approx   \frac{1}{2 r_{\mathrm{F}0}^2 } \left( x_j^2  - 2 x_j  V_x t  + x_j^2 \left( \frac{\Delta\nu}{\nu_0}\right) \right) 
\end{align}
Thus, the geometric phase increases linearly with time and with frequency.
As we discuss in Section\ \ref{sec:TypicalParameters}, for a few observations $V_x T/2r_\mathrm{F}$ be be as great as about 5;
and for some planned observations $\Delta\nu/\nu_0$ can approach 1. 
Section\ \ref{sec:WideBandLongTime} discusses the effects of the neglected terms in these cases.

\subsubsection{Effects of Internal Structures of Strips}\label{sec:InternalStructureUncertaintyPrinciple}

For observations with narrow bandwidth or short observing time, 
the structure factor $\Gamma_j$ will be nearly constant.
This is a consequence of the narrow width of strip, and the convolution in the definition of $\Gamma_i$, in Equation\ \ref{eq:GammaDef}.
That convolution smooths $\Gamma_j$ over a scale $W_q=4\pi/w$, in the domain of $q_j = (x_j - V_x t)/r_\mathrm{F}^2$.
The bandwidth $B$ or time $T$ required to resolve a particular structure of width $w$ is thus:
\begin{alignat}{2}
\frac{B}{\nu_0} &>\frac{4\pi r_\mathrm{F0}^2}{ x_j w} \\
V_x T &>\frac{4\pi r_\mathrm{F0}^2}{w}
\end{alignat}
We argue that strips of width greater than a few pairs of Fresnel zones are inefficient scatterers, so we might expect $w\lesssim 2\pi r_\mathrm{F}^2/x_j$.
To resolve this narrow width, the criteria above become:
\begin{alignat}{2}
B &\gtrsim  2 \nu_0 \\
V_x T &\gtrsim 2 x_j
\end{alignat}
Consequently, I assume that:
\begin{align}
\label{eq:GammaJApproxConst}
\Gamma_j \approx \mathrm{const.}
\end{align}
Long integration times might uncover effects of structure, particularly for features close to the apex with small $x_j$. 
Interestingly, Macquart (personal communication) has noted reduced coherence in frequency among sub-bands for observations of pulsar B0834+06, suggesting that some noodles may be rather wide. 
I discuss effects of internal structures of noodles quantitatively in Paper II.

\subsubsection{Approximated Kirchhoff Integral}

I apply the approximations of 
$x_j \gg r_\mathrm{F}$, $w \ll r_\mathrm{F}$, and $V_x t/w \ll r_\mathrm{F}$ (Sections\ \ref{sec:Approx:StripOffset}, \ref{sec:Approx:StripWidth}, and \ref{sec:Approx:TimeRange}) to write 
 the approximate form for the Kirchhoff integral in Equation\ \ref{eq:NetPsiFromStripToBeFourierTransformed}:
\begin{align}
\label{eq:Approx:KirchhoffIntegral}
\psi_\mathrm{obs} 
&= \psi_\mathrm{NS} + \Gamma_j e^{ i \phi_{\mathrm{g}j}} \psi_\mathrm{src}
\approx \psi_\mathrm{src}   + \frac{1}{\sqrt{2\pi} r_\mathrm{F}} e^{ i \phi_{\mathrm{g}j} - i \frac{\pi}{4}}  \left( \int_{-w/2}^ {+w/2} du\, e^{\frac {i}{r_\mathrm{F}^2} x_j u } \left[ e^{i\varphi_\mathrm{s} (u)}  - 1 \right] \right) \psi_\mathrm{src} 
.
\end{align}
where the geometric phase $\phi_{\mathrm{g}j}$ is approximated by Equation\ \ref{eq:ApproxPhiGeomDefFreqVariation}.
The quadratic factor of $u^2$ in the exponent of the integrand of Equation\ \ref{eq:DiffractiveStripSimplified} is omitted because $u\le w$, and $w\ll  r_\mathrm{F}, b, s$
as discussed in Section\ \ref{sec:Approx:StripWidth};
and the term $b_{1x} u$ is omitted from that integral because of the argument in Section\ \ref{sec:InternalStructureUncertaintyPrinciple}.
The result of the integral is on the order of $w$, so that the contribution of the strip is on the order of $w/r_\mathrm{F}$.

\subsection{Reference Model for Scintillation Arcs}\label{sec:ReferenceModel}

The approximations of Section\ \ref{sec:Approximations}, applied to the noodle model of Section\ \ref{sec:KirchhoffIntegral}, lead to scintillation arcs.
I apply the approximate form for $\phi_{\mathrm{g}j}$ as given by Equation\ \ref{eq:ApproxPhiGeomDefFreqVariation} and the assumption that $\Gamma_j\approx \mathrm{const}$ (Section\ \ref{sec:InternalStructureUncertaintyPrinciple}) to the expression for the observed field from multiple strips, Equation\ \ref{eq:MultiStripFieldGeneral}, to find:
\begin{align}
\label{eq:ReferenceModelTauF}
\psi_\mathrm{obs} 
&=  \psi_\mathrm{NS}   + \sum_j \left( \Gamma_j e^{i  \phi_{\mathrm{g}j}}  \right) \psi_\mathrm{src}
= \psi_\mathrm{NS}   + \sum_j \left(  \gamma_j e^{2\pi i  ( \alpha x_j^2 \Delta\nu  -\beta x_j t ) } \right) \psi_\mathrm{src}  
=\psi_\mathrm{NS}   + \sum_j \left(  \gamma_j e^{2\pi i  ( \tau_j \Delta\nu  -f_j t ) } \right) \psi_\mathrm{src}  
\end{align}
where I have defined
\begin{alignat}{3}
\label{eq:DefineAlphaBetaGamma}
\alpha &\equiv \frac{1}{4\pi r_{\mathrm{F}0}^2 \nu_0} = \frac{ (1+M) }{2 c D}, 
\qquad\qquad\qquad& 
\tau_j &\equiv \alpha x_j^2  \\
\beta &\equiv  -\frac{V_x}{2\pi r_{\mathrm{F}0}^2} =   -\frac{(1+M) }{c D} \nu_0 V_x ,  & f_j      &\equiv \beta x_j         \nonumber \\
\gamma_j &\equiv \Gamma_j e^{ {i x_j^2 }/{2 r_{\mathrm{F}0}^2} } 
.
\quad\quad \nonumber
\end{alignat}
Note that $\gamma_j$ is a complex variable that depends on $j$, and $\tau_j$ and $f_j$ are real variables that depend on $j$; whereas $\alpha$ and $\beta$ are parameters independent of $j$, although they depend on the details of the observation.  Note that $\beta$ depends on observing frequency $\nu_0$, but $\alpha$ does not. Because $x_j$ is large compared with other transverse dimensions, the geometric phase that contributes to $\gamma_j$ is more or less random.

The intensity is the square modulus of the field:
\begin{align}
I_\mathrm{obs} (\nu,t) &= \psi_\mathrm{obs} \psi_\mathrm{obs}^* \\
&\approx \left| \psi_\mathrm{src} \right|^2 \Bigg\{ \left[1 + \sum_j \left| \gamma_j \right|^2 \right]
+ 2 \sum_j \mathrm{Re}\left[ \gamma_j \exp\left( 2\pi i ( \tau_j \Delta\nu  -f_j t  )\right) \right] 
+ 2 \sum_{k<j}  \mathrm{Re}\left[ \gamma_j \gamma_k^* \exp\left( 2\pi i \left( ( \tau_j-\tau_k ) \Delta\nu  -( f_j-f_k )  t  \right)\right)\right]
\bigg\} \nonumber 
\end{align}
The Fourier transform the intensity to the delay-rate domain yields:
\begin{alignat}{3}
\label{eq:FourierTransformToDelayRate}
\tilde I_\mathrm{obs}(\tau,f) 
 & \rlap{$\displaystyle = \int_{-\infty}^\infty d\Delta\nu\, e^{2\pi \tau\Delta\nu}  \int_{-\infty}^\infty dt\, e^{2\pi ft}\,  I_\mathrm{obs} (\nu,t)$ }
\\
\label{eq:TildeIDelayRate}
&= \left| \psi_\mathrm{src} \right|^2 \Bigg\{
&& \delta(\tau)\delta(f) \\
&&+&\sum_j \Big[
\gamma_j \delta\left(\tau-\tau_j \right)\delta\left( f + f_j \right) &&+
\gamma_j^* \delta\left(\tau+\tau_j \right)\delta\left( f - f_j \right)
 \Big] \nonumber \\
&&+&\sum_{k<j} \Big[ 
\gamma_j \gamma_k^* \delta\left(\tau-( \tau_j-\tau_k )  \right)\delta\left( f + ( f_j-f_k ) \right) &&+
\gamma_j^* \gamma_k \delta\left(\tau+( \tau_j-\tau_k ) \right)\delta\left( f - ( f_j-f_k )  \right)
 \Big] \Bigg\} \nonumber 
\end{alignat}
where $\delta(...)$ is the Dirac delta-function, and I have omitted the correction $\sum_j |\gamma_j|^2$ from the first term.
The secondary spectrum  $\tilde C(\tau,f)$ is
the square modulus of the intensity in the delay-rate domain:
\begin{align}
\label{eq:TildeCDelayRate}
\tilde C(\tau,f) = \left| \tilde I(\tau,f) \right|^2
\end{align}
This secondary spectrum is an assemblage of delta-functions along parabolic arcs.
Figure\ \ref{fig:reference_arcs} shows the results in graphical form.
The first term in Equation\ \ref{eq:TildeIDelayRate} is simply a delta-function at the origin, shown as $a$ in the figure.
For some set of $x_i$ spanning a range including $x_i=0$, the second term sum sketches two parabolas with apexes at the origin;
at $\tau>0$ for the first term in square brackets, and at $\tau<0$ for the second term.
The figure shows these as $b$ and $c$ respectively.
Because I assume $|\Gamma | \approx w/r_\mathrm{F} \ll 1$, the parabolas are fainter than the delta function at the origin.
The equation for the parabolas is given by the curvature $a$:
\begin{align}
\label{eq:ReferenceModelTauFParabolicScaling}
\tau_j(f_j) &= \pm a x_j^2 = \pm \frac{\alpha}{\beta^2}f_j^2 
\end{align}
where Equation\ \ref{eq:DefineAlphaBetaGamma} defines the constants $\alpha$ and $\beta$.
Curvature of the parabola increases quadratically with observing frequency: $a = \frac{\alpha}{\beta^2}\propto \nu_0^{-2}$.
The third sum sketches a set of yet fainter parabolas with apexes at each point of those parabolas: opening toward $-\tau$ from apexes at $\tau>0$ for the first term in the square brackets, and toward $+\tau$ from apexes at $\tau<0$ for the second.
Figure\ \ref{fig:reference_arcs} shows typical examples as $d$ and $e$. 
This is precisely the form of the scintillation arcs.
Taking the square modulus of the secondary spectrum leaves the delta-functions intact, while removing any phases.

The arcs arise from the linear and quadratic dependences on $x_j$ 
in Equation\ \ref{eq:DefineAlphaBetaGamma}.
The scattering screen includes an assemblage of more or less random values $x_j$.
These random values
appear quadratically in the coefficient of $\Delta \nu$,
and linearly in the coefficient of $t$, in Equation\ \ref{eq:ReferenceModelTauF}.
The 2D Fourier transform to the delay-rate domain then converts these to parabolic arcs.

In practice, sampling in time and frequency, and the limited spans of each, convert the continuous Fourier transform of Equation\ \ref{eq:FourierTransformToDelayRate} to a discrete Fourier transform.
The result is the same combination of delta-functions, convolved with response functions in delay and rate, multiplied by the ``shah'' function \citep{bracewell2000fourier}. 
The response function in delay is the Fourier transform of the bandpass function of the instrument, and for a single sub-band usually resembles a sinc function, the Fourier transform of the observed band. 
However, significant phase and amplitude departures from that form are not unusual, particularly if multiple sub-bands are combined.
The response function in rate is usually nearly a sinc function, the Fourier transform of the time interval, although intensity variations of individual pulses can complicate it.
The shah function is an infinite series of evenly-space Dirac delta-functions, named after the Russian letter ``$\Sha$''. 
It is useful for relating continuous and discrete Fourier transforms \citep{bracewell2000fourier}. 

\section{Observed Parameters for Scintillation Arcs}\label{sec:TypicalParameters}

To frame my discussion, I present parameters for observations of pulsar B0834+06 by \citet{2010ApJ...708..232B},
and of pulsar J0437-4715 by \citet{Reardon2018PhDT}. 
\citeauthor{2010ApJ...708..232B} found well-defined arcs and rich secondary arclets via single-dish and interferometric observations of B0834+06.
Their work provides the best-constrained observations of a scintillation arc to date. 
\citeauthor{Reardon2018PhDT} found that pulsar J0437-4715 displayed scintillation arcs, and showed interesting behavior including variations of arc curvature with the orbital velocities of the binary pulsar and of the Earth. This pulsar is among the closest and brightest pulsars, so that it is likely to display among the smallest Fresnel scales and displacements $x_j$ for any pulsar that shows scintillation arcs.
The parameters of the observations of these two pulsars illuminate the assumptions of Sections\ \ref{sec:Approximations} and their application in Sections\ \ref{sec:Implications} and\ \ref{sec:ReferenceModel}. Table\ \ref{tab:Observed} summarizes the measured and inferred parameters.

\begin{table}
 \caption{Observed and inferred parameters for typical scintillation arcs.}
 \label{tab:Observed}
 \begin{tabular}{lccrcrr}
  \hline
Parameter & Symbol & Units & B0834+06 &  J0437-4715 & J0437-4715\\
  \hline
 Observing frequency & $\nu_0$ & MHz & 326. & 732. & 1400. \\
 Spectral bandwidth & $B$ & MHz & 8. & 64. & 300. \\
 Screen Distance & $D$ & pc & 226. &157. & 157. \\
 Magnification & $M$ & & 0.55 & 1.37 & 1.37 \\
 Fresnel scale & $ r_\mathrm{F}$ & cm & $8.1 \times 10^{10}$ & $3.6\times 10^{10}$ & $2.6\times 10^{10}$ \\
 Maximum delay & $\tau_\mathrm{max}$ & $\mu$s & 930. & 0.9 & 0.26 \\
 Minimum delay &  $\tau_\mathrm{min}$ & $\mu$s & 10. & 0.1 & 0.03 \\
 Maximum offset & $x_\mathrm{max}$ & cm & $1.6\times 10^{14}$ & $3.3\times 10^{12}$ & $1.8\times 10^{12}$ \\
 Minimum offset &  $x_\mathrm{min}$ & cm &  $1.6\times 10^{13}$ & $1.1\times 10^{12}$ & $6.1\times 10^{11}$\\
 Arc curvature & $a$ & $\mathrm{s}^{-3}$ & 0.52 & 0.003 to 0.004 & 0.010 to 0.015 \\
 Speed & $V_x$ & $\mathrm{km\ s}^{-1}$ & 110. & 125. & 125.  \\
 Observing time & $T$ & s & 6500. & 3600. & 42000. \\
Maximum displacement & $V_x T/2$  & cm & $3.6\times 10^{10}$ & $2.3\times 10^{10}$ &2.6$\times 10^{11}$ \\
Reference & & & \citet{2010ApJ...708..232B} & \citet{Reardon2018PhDT} & \citet{Reardon2018PhDT}\\
  \hline
 \end{tabular}
\end{table}

\citeauthor{2010ApJ...708..232B} observed B0834+06 in four frequency sub-bands  of bandwidth $B = 8\ \mathrm{MHz}$, over a frequency range of $310 \le \nu \le 342.5\ {\rm MHz}$.
I adopt a reference frequency of $\nu_0 = 326\ \mathrm{MHz}$.
Their most sensitive baseline, from Arecibo to the Green Bank Telescope,
had a projected length of $b\approx 2300\ {\rm km}$, far less than a Fresnel scale, as Table\ \ref{tab:Observed} shows.
Using a software correlator with 131072 channels, they obtained secondary spectra for individual sub-bands with resolution of 125\ ns in delay $\tau$ to a maximum delay of 2.05\ ms, and a resolution of 0.15\ mHz in rate $f$ over a width of $\pm 80\ {\rm mHz}$.
These parameters correspond to an effective observing bandwidth of $B=8\ {\rm MHz}$, or a single sub-band,
and an effective observing time interval of $T=6500\ \mathrm{s}$.
Arc curvature scales with frequency over the 8 sub-bands as $a\propto \nu_0^{-2}$, as expected.
They noted that positions of features remained the same to 10\% over the frequency range covered by their 8 sub-bands,
except for one feature at $\tau=1$\ ms which showed a small change with frequency. 
This feature is also displaced from the primary arc.
\citeauthor{2010ApJ...708..232B} concluded that a large-scale plasma gradient might displace this feature by refraction;
\citet{2016MNRAS.458.1289L}, 
and \citet{2018MNRAS.478..983S} 
found additional evidence for this.
Although this feature may be atypical, 
a feature at $\tau=0.93\ \mathrm{ms}$ seems to be completely typical,
so we adopt it as the maximum observed $\tau$ in Table\ \ref{tab:Observed}.

\citet{Reardon2018PhDT} found remarkable scintillation arcs in pulsar timing observations of J0437-4715.
He found arcs in two frequency bands, $\nu_0 = 732\ \mathrm{MHz}$ and $1400\ \mathrm{MHz}$;
he calls these the 40-cm and 20-cm bands, respectively, for their wavelengths.
Details of the observations varied with epoch, but typical parameters were a bandwidth of $B=64\ \mathrm{MHz}$ and scans of an hour in the 40-cm band, and a bandwidth of $B=300\ \mathrm{MHz}$ and multiple scans interpolated to cover several hours in the 20-cm band.
\citeauthor{Reardon2018PhDT} detected two arcs, both visible at both frequencies with the expected scaling of curvature with observing frequency, $a \propto \nu_0^{-2}$.
He did not observe secondary arclets.
Table\ \ref{tab:Observed} gives parameters for the stronger of these two arcs, the ``primary'' arc; parameters are similar for the weaker one.

We calculated the entries in Table\ \ref{tab:Observed} as follows.
Both \citeauthor{2010ApJ...708..232B} and \citeauthor{Reardon2018PhDT} report values for $D$ and $M$, or equivalent quantities, at their observing frequencies $\nu_0$.
Equation\ \ref{eq:FresnelScaleDef} then yields $r_{\mathrm{F}0}$.
\citeauthor{2010ApJ...708..232B} discuss the maximum delay $\tau_\mathrm{max}$, and the minimum delay of identifiable distinct features $\tau_\mathrm{min}$.
For \citeauthor{Reardon2018PhDT}, the frequency resolution of the dynamic spectrum forms an instrumental limit to the maximum observable delay,
as his Figure 4.1 shows; features probably persist to higher delays.
Also for \citeauthor{Reardon2018PhDT}, the table shows the minimum delay for which the arcs appear as distinct structures; individual features may not be separable as this small delay.
From the maximum and minimum delays we find the maximum and minimum offsets $x_\mathrm{max}$ and $x_\mathrm{min}$ using Equation\ \ref{eq:DefineAlphaBetaGamma}.
Both papers report values for the curvature, $a$.
Note that the arc curvature varies for J0437-4715 as $V_x$ changes with orbital phase of the Earth and pulsar;
we adopt \citeauthor{Reardon2018PhDT}'s smallest value of $a$, corresponding to the largest $V_x$, for the table.
We use the values for $a$, $D$, and $M$, and Equations\ \ref{eq:DefineAlphaBetaGamma} and \ref{eq:ReferenceModelTauFParabolicScaling} to find $V_x$.

As Table\ \ref{tab:Observed} shows, observations of B0834+06, and of $J0437-$715 at 40\ cm, are in accord with the assumptions stated in Section\ \ref{sec:Approximations}.
However, 
for observations of J0437-4715 at 20\ cm,
bandwidth $B$ approaches observing frequency $\nu_0$,
and $|V_x T|/2$ is $10\ r_\mathrm{F}$ and half of $x_\mathrm{min}$.
The bandwidth is great enough to blur the arc through its effect on curvature, $a\propto \nu^{-2}$;
this is the only effect of wide-bandwidth observations, 
as I discuss in Section\ \ref{sec:WideBandLongTime}.
As I also discuss there, long-time observations cause blurring only of secondary arclets, which are not detected in the observations of J0437-4715.

\section{Extensions of the Simple Noodle Model}\label{sec:Extensions}

\subsection{Sheets or Filaments Extending Along the Line of Sight}

\begin{figure}
\centering
\includegraphics[width=0.40\textwidth]{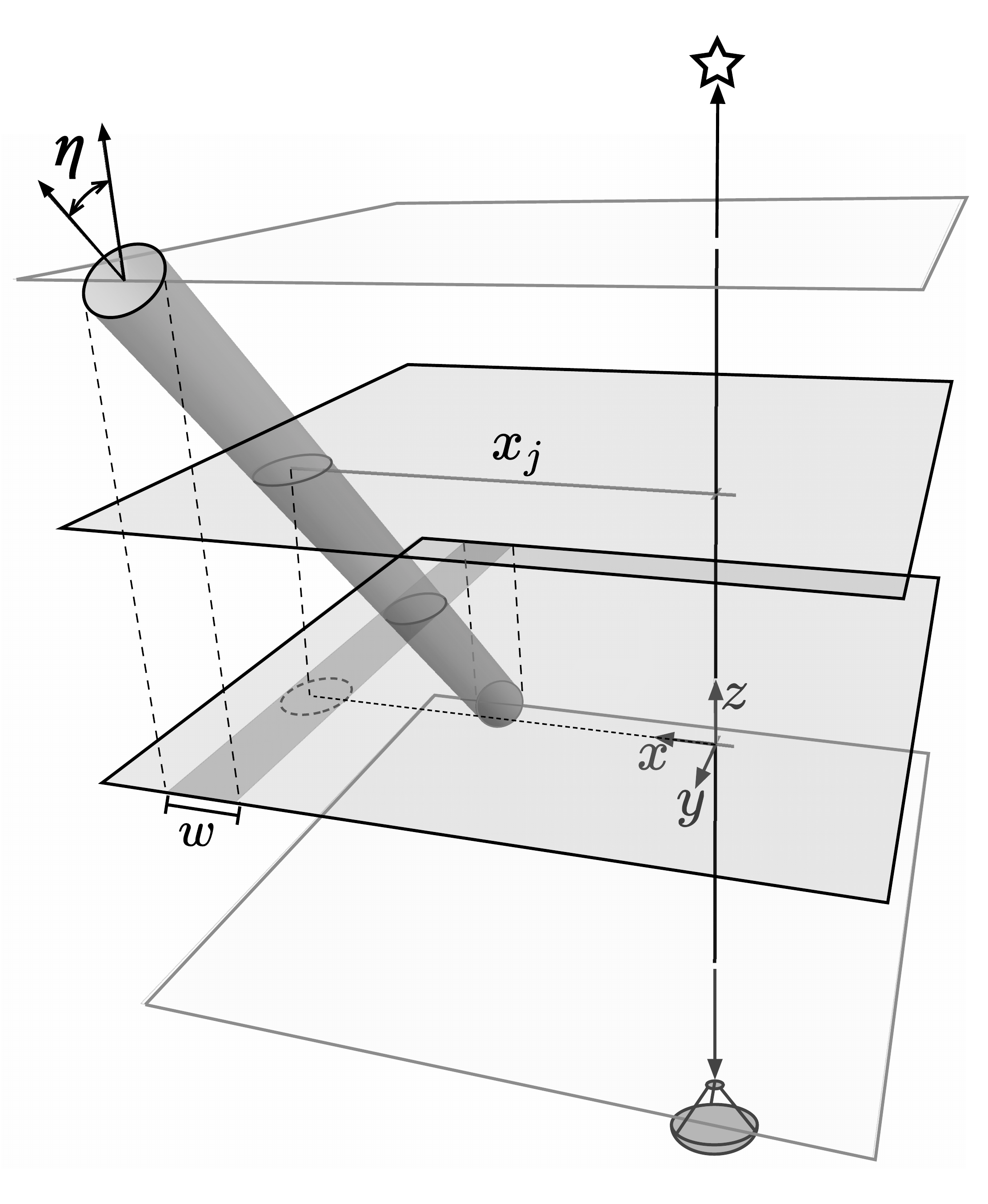}
\caption{A filament of scattering plasma tilted with respect to the line of sight acts as a strip of the same width, projected into a thin screen,
with separation $x_j$ equal to the closest approach of optical axis and filament.
The slant increases the effective column density of the filament.
The figure exaggerates the length and thickness of the filament, relative to the length of the line of sight.
\label{fig:TiltedSpaghetti}}
\end{figure}

Sheets or filaments of refracting material (``lasagna'' or ``spaghetti'') that extend along the line of sight, and are aligned with 2D surfaces of constant Fresnel phase,
can play the role as strips in a thin screen.
Figure\ \ref{fig:TiltedSpaghetti} shows the geometry for a filament.
Such sheets reduce the required plasma over- or under-density of a strip, by distributing the scattering plasma along the line of sight.
The effective column density is $\int N_e\csc\eta \, dz$, rather than $\int N_e \, dz$, where $\eta$ is the angle between the sheet or filament and the line of sight.
Because the cosecant is strongly peaked near $\eta=0$, this effect will strongly select for structures nearly parallel to the line of sight.
This is a fundamental part of the models of \citet{2014MNRAS.442.3338P}, \citet{2016MNRAS.458.1289L}, and \citet{2018MNRAS.478..983S}; in the model of \citet{2018MNRAS.478..983S} a radius of curvature plays the role of $w\csc\eta$.

Sheets or filaments that lie on surfaces of constant Fresnel phase scatter coherently, and so act as if they were projected onto the Fresnel screen.
Such surfaces take the form of ellipsoids of rotation about the undeflected line of sight, with the foci of the ellipsoids at the source and observer positions.
Curvature of the surface, or of the sheet or filament, limits their extent along the line of sight. 
In the approximations considered in Section\ \ref{sec:Approximations},
the possible distance of such extension is quite long, although the sheet or filament must be quite thin:
the Fresnel phase changes by 1 radian in a distance parallel to the optical axis of $\Delta z \approx r_\mathrm{F} \sqrt{ (R+D)/ x_j}$,
as compared with $r_\mathrm{F}^2/x_j$ perpendicular to it.
For pulsar B0834+06 as discussed in Section\ \ref{sec:TypicalParameters}, $\Delta z \approx 4\times 10^{14}\ \mathrm{cm}$ along the line of sight for a sheet at $x_j=10^3\ r_\mathrm{F}$, with a thickness of $w\lesssim 2\pi r_\mathrm{F}^2/x_j \approx 3\times 10^8\ \mathrm{cm}$.
Evidently the physics of such structures limits their extent along the line of sight.

\subsection{Canted Noodles}\label{sec:CantedNoodles}

\begin{figure}
\centering
\includegraphics[width=0.30\textwidth]{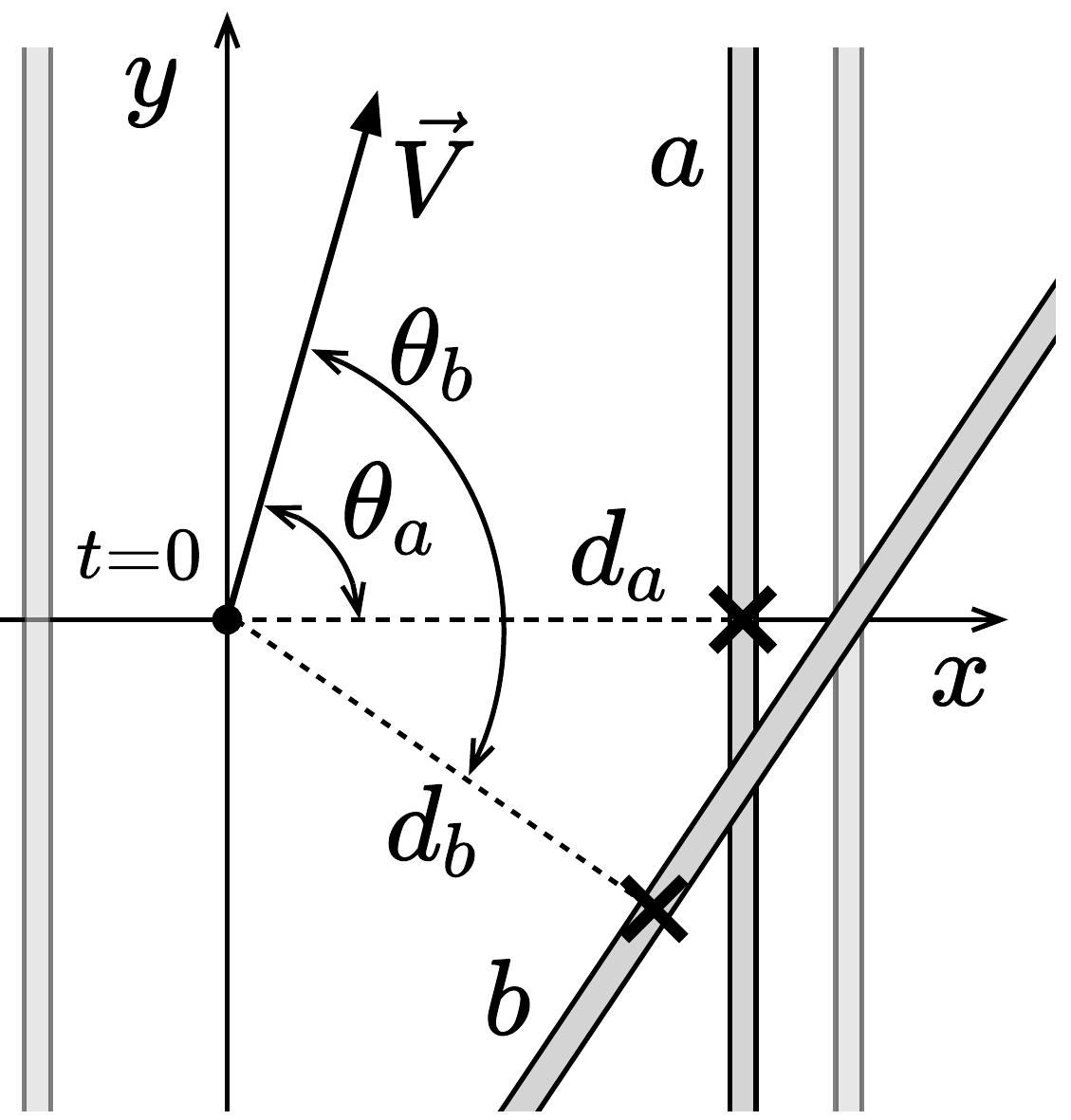}
\caption{Schematic view of scattering by a parallel strip (a) and a canted strip (b). 
A canted strip will combine with the undeflected line of sight to produce an point off the parabolic arc defined by the other, parallel strips;
and with the other strips to produce an arclet there.
Crosses show the points where strips are closest to the undeflected line of sight at time $t=0$.
The thickness of the strips are highly exaggerated.
\label{fig:BentStrip}}
\end{figure}

If one strip lies at an angle to the others, then it will produce a point in the secondary spectrum that lies off the original arc.
Figure\ \ref{fig:BentStrip} shows an exaggerated view of the geometry.
Mathematically, the offset of the resulting point $(\tau_j,f_j)$ reflects the inclination $\theta_b$ for this ``canted'' strip,
resulting in a different value for $\beta$ in Equation\ \ref{eq:DefineAlphaBetaGamma} for this particular strip.
We define $d_j$ as the distance of the line of sight from the strip at its closest point, $j$ at $t=0$. For strips parallel to the $y$-axis, $d_j=x_j$.
We define $\theta_j$ as the angle between strip $j$ and the velocity $\mathbf{V}_1 = \frac{d}{dt}\mathbf{b}_1$ and the normal to the strip.
The component of velocity parallel to the strip is $V_1 \cos\theta_j$.
We then revise the definitions in\ \ref{eq:DefineAlphaBetaGamma}:
\begin{alignat}{3}
\alpha &\equiv \frac{ (1+M) }{2 c D},                      
\qquad\qquad\qquad& 
\tau_j &\equiv \alpha x_j^2  \\
\beta_j & \equiv  -\frac{(1+M) }{c D} \nu_0 V_1 \cos\theta_j  ,  & f_j      &\equiv \beta_j d_j         \nonumber \\
\gamma_j &\equiv \Gamma_j e^{ {i x_j^2 }/{2 r_{\mathrm{F}0}^2} } 
.
\quad\quad \nonumber
\end{alignat}
If the screen contains two cohorts of strips with different directions $\theta_a$, $\theta_b$, then two main arcs will be observed,
with different curvatures $a$, $b$.
Interestingly, if secondary arclets are present, they should appear on both arcs with both curvatures.
If we suppose that one strip is canted at $\theta_c$, and the rest are parallel, then the point in the secondary spectrum from 
interference of that strip and the undeflected line of sight will lie off the main parabola;
its rate will differ from that of the main parabola at the same delay by a factor of $\cos\theta_c/\cos\theta_\parallel$.
The secondary arclet extending from this point will have the same curvature as the primary arc;
and the canted strip will contribute an offset point to the secondary arclets from other strips.
Although this could explain the offset of the 1-ms feature observed by \citet{2010ApJ...708..232B}, evidence seems to favor displacement by a refracting lens or prism \citep{2016MNRAS.458.1289L,2018MNRAS.478..983S}.

The populations of canted and parallel noodles, and their directions, depend on the physical nature of the noodles and their environment. If the noodles are magnetic field lines (or bundles of field lines) in a reconnection sheet, then one expects at least two directions for the field on large scales, although both might not carry plasma density fluctuations.  The multiple arcs seen for most pulsars with scintillation arcs might reflect multiple field directions. Comparison with theoretical calculations of field geometry in reconnection sheets, as well as additional observations of arcs, may improve understanding of both reconnection and arcs.

\subsection{Curved, Bent, and Finite-Length Noodles}

If noodles curve over a scale longer than the Fresnel scale, then the stationary-phase approximation holds for the integral along the noodle, as over $y$ in Equation\ \ref{eq:DiffractiveStrip}. As long as the region of stationary phase has dimension of about $r_\mathrm{F}$, all results are unchanged.
The same holds true if the width, plasma column, or other properties change over scales larger than $r_\mathrm{F}$.

If a noodle has finite length, and with a sharp cutoff relative to the Fresnel scale, then for a straight noodle the integral over $y$ yields the complex error function.\ \citep[See, for example][Ch. 7.]{abramowitz2012handbook} The result is the same as that from Equation\ \ref{eq:DiffractiveStrip}, with an additional oscillating correction that falls in inverse proportion to the distance from the point where the noodle passes closest to the optical axis, $\mathbf{d}_j$ to the end of the noodle.
Similarly, if the noodle has a sharp bend, the result of integration would be a sum of two complex error functions, resulting in an oscillating correction term.
One expects that geometrically sharp changes would be rare, because of magnetic tension along magnetic field lines and the ease of plasma transport along them, as discussed in Section\ \ref{sec:PhysicalNature}. A sharp bend would travel along the noodle at the Alfv\'en speed, about $20\ \mathrm{km\ s}^{-1}$ \citep{1990ApJ...353L..29S}.

\subsection{Time Variations of Noodle Structure}\label{sec:TimeVariations}

The region of scattering is so small, and noodles are so narrow, that their shapes might change within an observation.
At the Alfv\'en speed of about $20\ \mathrm{km\ s}^{-1}$ \citep{1990ApJ...353L..29S}, a disturbance can cross a pair of Fresnel zones at $1000 r_\mathrm{F}$ for pulsar B0834+06 at $\nu_0=326\ \mathrm{MHz}$ in about 40\ s, and the Fresnel scale in about 10\ hr.
So, changes between observations are expected, and changes local to a strip within a single observation would not be surprising. 
The observed stability of arc parameters over long periods suggests that the global geometry of magnetic fields is uniform over thousands of AU.
Similar considerations hold for pulsar J0437-4715 \citep{Reardon2018PhDT}.

\subsection{Wide-Bandwidth and Long-Time Observations}\label{sec:WideBandLongTime}

\begin{figure}
\centering
\includegraphics[width=\textwidth]{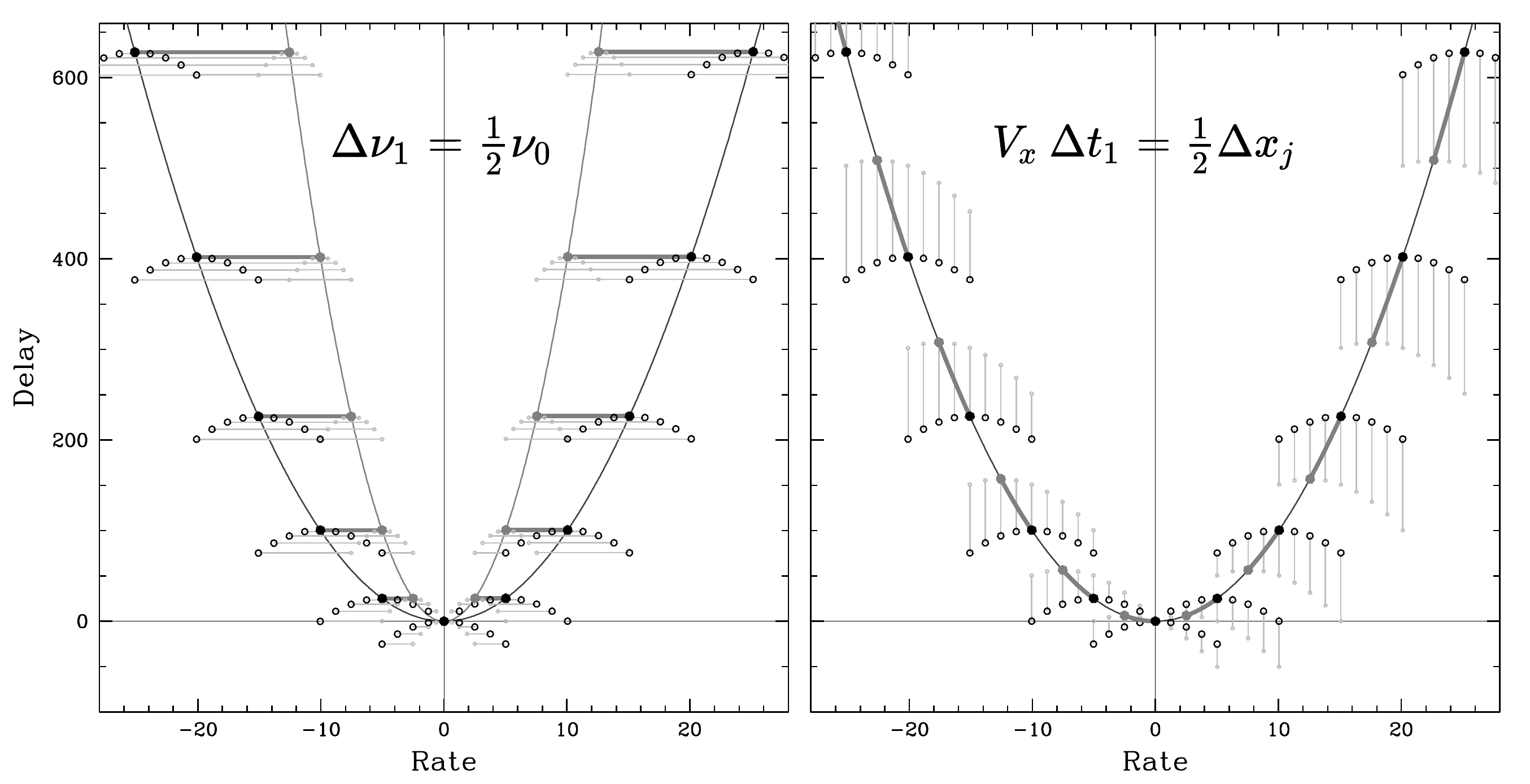}
\caption{Effects of observations with wide bandwidth (left panel) and over long times (right).
At left, a decrease of observing frequency by a factor of 2 moves all points from the narrower arc and accompanying arclets (lighter gray) to the broader arc and its arclets (black). 
Observing over the entire bandwidth smears points along the indicated line segments.
At right, observing by a time equal to the spacing of indicated points, divided by the transverse speed $V_x$, shifts points on the primary arc along the arc, and points on the secondary arclet in rate only, from gray to black as time increases.
A single scan over the entire time will smear points along line segments.
\label{fig:WideBandwidthLongTime}}
\end{figure}

Coherent observations over wide bandwidths, or over long times, will tend to smear features in the secondary spectrum.
The most prominent such effect is that wide bandwidth smears 
point features in the secondary spectrum into line segments parallel to the rate axis $f$,
because arc curvature depends on observing frequency, as Equations\ \ref{eq:DefineAlphaBetaGamma}\ and\ \ref{eq:ReferenceModelTauFParabolicScaling} show.
Figure\ \ref{fig:WideBandwidthLongTime} illustrates the effect.
On the other hand, very long coherent integration time smears point features into line segments along the arc,
and translates arclets in delay, as Figure\ \ref{fig:WideBandwidthLongTime} shows.
In effect, changes in delay or rate of the scattered paths from wide bandwidth or long integration time will reduce coherence in frequency or time, respectively, so that the result is the incoherent sum of the observations with narrower bandwidth or shorter time.

I now consider the effects of wide bandwidth and long time quantitatively.
I consider the effects on the geometric phase, and consequent changes of the secondary spectrum.
I do not include effects of source structure, as discussed briefly in Sections\ \ref{sec:InternalStructureUncertaintyPrinciple} and\ \ref{sec:TimeVariations} above; quantitative discussion requires a model for source structure, so I leave it to Paper II.

Equation\ \ref{eq:MultiStripIntensityGeneral} shows that only the differences of the geometric phase affect observables.
These differences are: 
\begin{alignat}{2}
\phi_{\mathrm{g}j}-\phi_\mathrm{NS} &=
\frac{1}{2 r_\mathrm{F}^2 }  \left( x_j - b_{1x} \right)^2 
&&= \frac{\pi (1+M) \nu }{c D} \left( x_j -V_{x} t \right)^2 
\\
\intertext{for the terms responsible for the primary arc; and }
\phi_{\mathrm{g}j}-\phi_{\mathrm{g}k} &=\frac{1}{2  r_\mathrm{F}^2 }  \left[ \left( x_j^2 -  2 b_{1x}  x_j\right) - \left( x_k^2 - 2 b_{1x} x_k\right)\right]
&&= \frac{\pi (1+M) \nu }{c D} \left( (x_j^2 -x_k^2) - 2 (x_j-x_k) V_{x} t \right)
\end{alignat}
for those responsible for the secondary arclets.
Here again, $b_{1x}=(b_x +M s_x)/(1+M)\equiv V_x t$.
We explicitly include dependence on frequency and time, and assume that source and observer are on the optical axis at $t=0$.
Note that the phase difference for the primary arc contains a term proportional to $b_{1x}^2$,
and the difference for the secondary arclet does not. This is because the undeflected line of sight is free to move in $x$, whereas the deflected lines of sight are not.
The undeflected line of sight has the same character as a canted strip that passes through the optical axis and is parallel to the velocity $\mathbf{V}$,
through second order in $\mathbf{b}_1$.

The change of geometric phase over narrow ranges of frequency and time is nearly linear, so that each phase difference leads to a point in the delay-rate domain. For narrow ranges centered at $\nu_1$ and $t_1$, the locations of those points are:
\begin{alignat}{2}
\label{eq:TaugjSRategjS}
\tau_j &= \frac{1}{2\pi}\left. \frac{\partial  ( \phi_{\mathrm{g}j}-\phi_\mathrm{NS} )}{\partial \nu} \right|_{\nu_1,t_1} \!\!  =  \frac{(1+M) }{2 c D} \left( x_j -V_{x} t_1 \right)^2 ,
\quad\quad\quad\quad\quad\quad &
f_j &= \frac{1}{2\pi}\left. \frac{\partial ( \phi_{\mathrm{g}j}-\phi_\mathrm{NS}) }{\partial t}\right|_{\nu_1,t_1} \!\! =  -\frac{(1+M) }{c D} V_x \left( x_j -V_{x} t_1 \right) \nu_1 
\\
\intertext{for the primary arc; and }
\label{eq:TaugjkRategjk}
\tau_{jk} &=\frac{1}{2\pi} \left. \frac{\partial ( \phi_{\mathrm{g}j}-\phi_{\mathrm{g}k} ) }{\partial \nu}\right|_{\nu_1,t_1} \!\!  =  \frac{(1+M) }{2 c D} \left( (x_j^2 -x_k^2) - 2(x_j-x_k) V_{x} t \right),
  &
f_{jk} &= \frac{1}{2\pi} \left. \frac{\partial ( \phi_{\mathrm{g}j}-\phi_{\mathrm{g}k} ) }{\partial t}\right|_{\nu_1,t_1} \!\!  = - \frac{ (1+M) }{c D}  V_x   (x_j-x_k) \nu_1
\end{alignat}
for the secondary arclets.
Integration over larger ranges of frequency smears the points over $\nu_1$ and over longer times smears them over $t_1$.
This smearing is a consequence of the stationary-phase approximation, applied to the Fourier transform that relates the frequency-time and delay-rate domains, Equation\ \ref{eq:FourierTransformToDelayRate}.

Inspection of Equations\ \ref{eq:TaugjSRategjS} and\ \ref{eq:TaugjkRategjk} shows that rate $f$ depends on frequency, but delay $\tau$ does not, for both primary and secondary arclets. Indeed, rate is proportional to frequency $\nu_1$ for both. Thus, integration over wide bandwidth smears the secondary spectrum in rate, horizontally in Figure\ \ref{fig:reference_arcs}. This this is another manifestation of the fact that arc curvature depends on frequency through the frequency dependence of $\beta$, as discussed in Section\ \ref{sec:ReferenceModel} above, following Equations\ \ref{eq:DefineAlphaBetaGamma} and\ \ref{eq:ReferenceModelTauFParabolicScaling}. This feature of the noodle model is in accord with observations\ \citep{2003ApJ...599..457H}.

Integration over time has slightly different effects on primary and secondary arclets, because of the extra quadratic dependence on $b_{1x}$ for the primary arc in Equation\ \ref{eq:TaugjSRategjS}.
For the primary arc, the quadratic dependence of $\tau_j$ on $t$ and linear dependence of $f_j$, and their coefficients, means that a point on the primary arc will remain on the arc, but move along the arc as time progresses. This behavior is well-observed in multi-epoch observations: features tend to move along the arc with time. For long single observations, this leads to smearing along the arc.
For the secondary arclet, only delay depends on time, and its dependence is linear. Individual points on a secondary arclet will remain at the same rate, but move up or down in delay, at a rate proportional to the separation of strips, $x_j - x_k$. In consequence, points on the arclet move up or down in Figure\ \ref{fig:reference_arcs} by different amounts, in such a way that the arclet maintains the same curvature, but the apex shifts among points so that the apex remains on the primary arc. 
As the arclet shifts, selection effects will maintain the strongest part of the arclet on the arc, discussed in Section\ \ref{sec:OpticalDepth} and Paper II.

A critical element in analyzing wideband data for scintillation arcs is calibration of the frequency response.
Ideally, the response would have unit amplitude and constant phase over the observed bandwidth, and zero amplitude outside it.
In practice, amplitude and phase both vary with frequency.
For single-dish observations in a single polarization, phase variations are unimportant, but the secondary spectrum of the source is convolved with the square modulus of the Fourier transform of the passband.
This can complicate analysis.
Usually, wideband observations are accomplished by dividing the observed band into many sub-bands, each with its own amplitude and phase profiles.
Correcting for these effects can be difficult. \citet{2010ApJ...708..232B}, 
for example, analyze only one of their four sub-bands.
Recently has it become possible to digitize and Fourier transform wider bandwidths in a single sub-band, so that wideband observations may become easier and more useful.

\subsection{Interferometric Observations}\label{sec:Interferometric}

Interferometric observations measure the locations of the centers of the strips in $x$; and of the stationary-phase points in $y$, where the noodles approach the undeflected line of sight most closely.
Demonstrating this fact involves calculating the interferometric visibility,
analogously to the calculation of intensity in Equation\ \ref{eq:MultiStripIntensityGeneral}, but using the product of the observed fields $\psi_\mathrm{obs}(\mathbf{b})$ at two different locations in the observer plane $\mathbf{b}_\mathrm{A}$, $\mathbf{b}_\mathrm{B}$, as given by Equation\ \ref{eq:psi_obs_GeneralForm}.
The calculation is not difficult, but requires defining new quantities and writing some long equations, so we summarize it here.
For interferometer baseline lengths of less than about $r_\mathrm{F}$, as is the case for all existing and planned interferometers,
both locations for antennas in the observer plane lead to the same structural amplitude and phase from the strip $\Gamma_j$in Equation\ \ref{eq:PhasorForOneStrip}. However, they have a different geometric phase $\phi_{\mathrm{g}j}$ that reflects the difference in positions of antennas in the observer plane, $\mathbf{b}_\mathrm{A}$ and $\mathbf{b}_\mathrm{B}$.
The phase differences that arise in calculation of the visibility $V= \psi_\mathrm{obs}(\mathbf{b}_\mathrm{A}) \psi_\mathrm{obs}(\mathbf{b}_\mathrm{B})$
are simply the differences of this geometric phase.
The observed angular offset for the apex of an arclet is then simply $x_j/D$ along the $x$-direction and $0$ along the $y$-direction, as expected from Equation\ \ref{eq:MultiStripIntensityGeneral}. Equation\ \ref{eq:DefineAlphaBetaGamma} gives the delay and rate at the apex of the arcle. The angular offset between points on the arclet is  $(x_j-x_k)/D$. 
Thus, strictly parallel noodles give rise to a strictly linear arrangement of subimages, as inferred from interferometry. A canted noodle gives rise to a subimage off the axis of that structure, by the angle $\theta_\mathrm{a}-\theta_\mathrm{b}$ in Figure\ \ref{fig:BentStrip} and Section\ \ref{sec:CantedNoodles}.

\section{Summary}\label{sec:Summary}

I use Kirchhoff diffraction to show a simple model for scintillation arcs, based on ``noodles'' of over- or under-dense plasma, generically produces scintillation arcs. The noodles are much longer than they are wide: they may take the form of filaments or sheets. Mathematically, their effect is that of strips projected into a thin screen perpendicular to the line of sight. These strips extend over many Fresnel zones along their lengths, but have width of about a pair of Fresnel zones perpendicular to their lengths, and to the line of sight. 
At the inferred separation of the strips from the undeflected line of sight, this requirement is for lengths of hundreds or thousands of times their widths.
The noodles plausibly follow magnetic field lines; perhaps in reconnection sheets, where strong turbulence is expected on small scales.
Observations of scintillation arcs would then allow visualization of fields in reconnection regions.
I make the assumption that the phase change across the strip $\varphi_\mathrm{s}$ does not depend on position along it; this is plausible because charged particles move easily along field lines, and with difficulty across them.
The Kirchhoff integral along the strip is then straightforward integration of a complex Gaussian function;
the integral across the strip is the Fourier transform of $\varphi_\mathrm{s}$ minus the contribution of the strip with $\varphi\equiv 0$.
The main arc arises from interference between a strip and the undeflected line of sight; or, equivalently, between a strip and a population of wide strips that scatter only at small angles.
The secondary arclets arise from interference among strips. 
If the number of strips is few, so that most radiation is not scattered, then the scintillation arcs are weak relative to the undeflected line of sight, and the arclets are weaker still, as observed.
Because an arc scatters most efficiently when its width is about that of a pair of Fresnel zones or less, and the magnitude of the scattered field is proportional to its width, the main arc will be brightest near the apex, and arclets will be brightest at their apexes on the primary arc. 
This selection effect will ensure that a ``cloud'' of scattering noodles always lies near the undeflected line of sight.
As time progresses, arclets and other features will move along the primary arc, toward increasing rate.
Strips that are not parallel, or ``canted,'' relative to the rest, will produce arclets off the main arc, and corresponding features separated from the secondary arclets of other strips.
Cohorts of strips parallel to different directions, as expected in regions of reconnection, will lead to multiple arcs.
I discussed the effects of curved, bent, and finite-length noodles;
and the results expected from interferometric observations.

\section*{Acknowledgements}

I thank Dana Simard, Ue-Li Pen, and Dan Stinebring for useful conversations, and Walter Brisken for sharing his data.
I thank the T.D. Lee Institute of Shanghai Jiao Tong University for hospitality during an essential phase of this work.

%
%
%
\bibliographystyle{mnras}
\bibliography{extinctbib} 



\bsp	
\label{lastpage}
\end{document}